\documentclass[sigconf]{acmart}
\usepackage{subfigure}

\usepackage[ruled,linesnumbered]{algorithm2e}
\AtBeginDocument{%
  \providecommand\BibTeX{{%
    \normalfont B\kern-0.5em{\scshape i\kern-0.25em b}\kern-0.8em\TeX}}}

\setcopyright{iw3c2w3}

\begin{document}
\copyrightyear{2021}
\acmYear{2021}
\acmConference[WWW '21]{Proceedings of the Web Conference 2021}{April 19--23,
2021}{Ljubljana, Slovenia}
\acmBooktitle{Proceedings of the Web Conference 2021 (WWW '21), April 19--23, 2021,
Ljubljana, Slovenia}
\acmPrice{}
\acmDOI{10.1145/3442381.3449873}
\acmISBN{978-1-4503-8312-7/21/04}

\title{Adversarial and Contrastive Variational Autoencoder for Sequential Recommendation}

\author{Zhe Xie}
\email{xiezhe20001128@sjtu.edu.cn}
\affiliation{Department of Computer Science and Engineering\\Shanghai Jiao Tong University}
\authornote{Zhe Xie and Chengxuan Liu contribute equally.}

\author{Chengxuan Liu}
\email{sjtulcx@sjtu.edu.cn}
\affiliation{School of Cyber Science and Engineering\\Shanghai Jiao Tong University}
\authornotemark[1]

\author{Yichi Zhang}
\email{zhangyichi2017@sjtu.edu.cn}
\affiliation{University of Michigan-Shanghai Jiao Tong University Joint Institute\\Shanghai Jiao Tong University}

\author{Hongtao Lu}
\email{lu-ht@cs.sjtu.edu.cn}
\affiliation{Department of Computer Science and Engineering\\Shanghai Jiao Tong University}

\author{Dong Wang}
\email{wangdong@sjtu.edu.cn}
\affiliation{School of Software\\Shanghai Jiao Tong University}
\authornotemark[2]

\author{Yue Ding}
\email{dingyue@sjtu.edu.cn}
\affiliation{School of Software\\Shanghai Jiao Tong University}
\authornote{Dong Wang and Yue Ding are corresponding authors.}

\renewcommand{\shortauthors}{Zhe Xie, Chengxuan Liu, Yichi Zhang, Hongtao Lu, Dong Wang, and Yue Ding}

\begin{abstract}
 Sequential recommendation as an emerging topic has attracted increasing attention due to its important practical significance. Models based on deep learning and attention mechanism have achieved good performance in sequential recommendation. Recently, the generative models based on Variational Autoencoder (VAE) have shown the unique advantage in collaborative filtering. In particular, the sequential VAE model as a recurrent version of VAE can effectively capture temporal dependencies among items in user sequence and perform sequential recommendation. However, VAE-based models suffer from a common limitation that the representational ability of the obtained approximate posterior distribution is limited, resulting in lower quality of generated samples. This is especially true for generating sequences. To solve the above problem, in this work, we propose a novel method called Adversarial and Contrastive Variational Autoencoder (ACVAE) for sequential recommendation. Specifically, we first introduce the adversarial training for sequence generation under the Adversarial Variational Bayes (AVB) framework, which enables our model to generate high-quality latent variables. Then, we employ the contrastive loss. The latent variables will be able to learn more personalized and salient characteristics by minimizing the contrastive loss. Besides, when encoding the sequence, we apply a recurrent and convolutional structure to capture global and local relationships in the sequence. Finally, we conduct extensive experiments on four real-world datasets. The experimental results show that our proposed ACVAE model outperforms other state-of-the-art methods.
\end{abstract}

\ccsdesc[500]{Information systems~Recommender systems}

\keywords{Sequential Recommendation, Variational Autoencoder, Adversarial Learning, Contrastive Learning}


\maketitle

\section{Introduction}
With the rapid development of web technology, the amount of data is growing explosively, and we have to face massive information every day. The recommender systems (RSs) as an important tool to alleviate information overload can generate a personalized recommendation list for different users.

The core recommendation method in RSs is collaborative filtering (CF). Matrix decomposition \cite{MF} is the most representative approach in CF, which decomposes the observed matrix into two low-rank matrices of users and items.
Temporal dynamic as one of the classic problems in RSs has always attracted attention. Traditional CF methods such as matrix decomposition calculates user's temporal dynamic interest by introducing time segments, but it usually leads to inaccurate prediction results because it fails to capture the changes in dependencies between items over time. The appearance of sequential recommender systems (SRSs) brings a significant improvement in alleviating this problem. SRSs aim to find potential patterns and the dependencies of items in the sequence, and understand user's time-varying interest behind the sequence of his interacted items in order to accurately make next-item recommendation. Different from conventional RSs, SRSs consider the sequential context by taking the prior sequential interactions into account, which effectively models users' dynamic preference and items' popularity \cite{2019SRSs}.
 
In order to process the sequence information for SRSs, lots of methods are proposed to capture the sequential patterns. The Factorizing Personalized Markov Chain (FPMC) \cite{FPMC} uses Markov chain to obtain the information of previous items. Although FPMC is able to make recommendations using sequence information, long-term sequence information can not be captured due to the constraints of Markov chain. Because of the powerful learning ability of neural networks, recurrent neural networks (RNN) based method such as LSTM \cite{Wu2017RecurrentRN} can be utilized in capturing long-term information, which greatly improves the performance of SRSs. Besides, attention mechanism is widely used in recent models such as SASRec \cite{SASRec} and BERT4Rec \cite{BERT4Rec}. Compared with RNN-based models, attention-based models achieve better prediction results because of its ability in capturing global dependencies.

Generative Adversarial Network (GAN) \cite{CFGAN} and Variational Autoenoder (VAE) \cite{liang2018variational} as two powerful generative models have been successfully applied in RSs. Based on VAE, the sequential VAE (SVAE) \cite{SVAE} model employs a recursive implementation of standard VAE encoder to capture temporal features of items. SVAE has shown good predicting results with the powerful capability of sequence reconstruction. With a sequence of items as the input data, the model can generate "next-k" items that are most likely to be chosen by users. However, we argue that the encoder of SVAE has significant limitations, resulting in obtaining low quality of approximate posterior distribution. The deep learning information bottleneck theory reveals that the essence of deep learning is to eliminate useless information and leave real effective information \cite{alemi2016deep}. When applying VAE to generate sequences, we hope to get high-quality latent variables containing sufficient information so that the decoder can use it to generate high-quality samples. Besides, we hope that the latent variables sampled for different users can reflect obvious differences in order to construct "personalized" user sequences.

For the above motivation, in this paper, we propose Adversarial and Contrastive Variational Autoencoder (ACVAE), which has made several improvements to VAE model for sequential recommendation. This new model tries to reduce the unnecessary dependency constraints on latent variables and allow the model to learn more personalized and effective user characteristics, which can be reflected in two aspects: 1) different dimensions in the latent variables, 2) different latent variables between different users. Within one latent variable, the different dimensions should have low relevance. In this way, the information contained in each dimension in the latent variable will be unique and more representative. Between different users, their corresponding latent variables should have a certain difference, so that the latent variables will have more personalized and salient information. In this work, we find that Adversarial Variational Bayes (AVB) plays an important role in reducing the relevance of various dimensions of latent variables. Thus we introduce AVB into sequential recommendation, which brings the latent variables a more flexible approximate posterior distribution and enhances the independence of different dimensions. Then, we introduce a contrastive learning method for VAE model to assist the training of the encoder, which brings more personalized and salient characteristics of users to the latent variables. Finally, for the task of sequential recommendation, we add a special convolutional layer which can improve the RNN-based encoder in capturing local information between adjacent items. This enables the encoder to learn effective local dependencies in the input sequences. The main contributions of our work are as follows:
\begin{itemize}
    \item We propose to introduce adversarial learning under the AVB framework for sequential recommendation, which brings a closer approximate posterior distribution to the true distribution for sequential VAE model and reduces the correlation between different dimensions of latent variables.
	\item We introduce contrastive learning to VAE model and utilize contrastive loss to learn the distinctive users' characteristics. The optimization of contrastive loss ensures the personalized and salient characteristics of different users.
	\item We leverage a convolutional layer to strengthen the connections between adjacent items in the inference model, which helps to better capture short-term relationships in the sequence.
\end{itemize}

\section{Related Work}
\subsection{Sequential Recommendation}
Sequential recommendation performs next-item prediction according to users' historical interactions. Conventional collaborative filtering methods for recommendation usually fail to capture the dependencies of items in the sequence. Therefore, they are not suitable for sequential recommendation scenarios. Caser \cite{Caser} firstly uses a vertical and a horizontal Convolutional Neural Network (CNN) to capture the local sequence information. In order to describe users' dynamic preference and items’ popularity over time, FPMC \cite{FPMC} introduces Markov chains to capture the dependency of the previous item. Following FPMC, higher-order Markov chain \cite{HighOrderMC} is used in learning high-order sequential dependencies. Besides, some models that learn long-term sequential dependencies use LSTM \cite{Wu2017RecurrentRN} and GRU \cite{hidasi2015sessionbased}. Hierarchical structures are also used for improving the performance of sequential recommendation. For example, Parallel RNN \cite{FeatureEnrichedRecommendation} brings both the user-item interactions and meta data together. RCNN \cite{RCNN} combines recurrent layer and convolutional layer together to mine short-term sequential patterns. Attention mechanism is a popular technology recently. SASRec \cite{SASRec} brings self-attention into SRSs and BERT4Rec \cite{BERT4Rec} employs BERT model to learn bidirectional item dependencies for sequential recommendation. Hierarchical attention network \cite{HAttn} is used for capturing both long-term and short-term sequence information. 

\subsection{VAE for Recommendation}
Variational Autoencoder \cite{kingma2013auto,rezende2014stochastic} learns the approximate posterior of latent variables under the variational inference framework. There are also some variants of VAE, such as $\beta$-VAE \cite{higgins2016beta} which learns disentangled representations and DVAE \cite{im2017denoising} which is similar to denoising autoencoder. The encoded user preference variables (\emph{i.e.}, the latent variables) in variational autoencoder can be used in generating the distribution of recommended items. Mult-VAE \cite{liang2018variational} is a representative method of using VAE for recommendation. Based on Mult-VAE, SVAE \cite{SVAE} takes in a sequence of items with sequential dependencies, and processes it with the GRU network and finally outputs probability distribution of candidate items. CVRCF \cite{CVRCF} employs a recurrent neural network and includes both user and item features in variational inference. MacridVAE \cite{MacridVAE} employs VAE to learn disentangled representations that can enhance robustness. RecVAE \cite{RecVAE} proposes a new composite prior for training based on alternating updates to enhance performance.

\subsection{Adversarial Learning}
Adversarial learning has been successfully utilized in some models like APR \cite{APR}. GAN-based models such as IRGAN \cite{IRGAN}, CFGAN \cite{CFGAN} can be used for recommendation. SeqGAN \cite{yu2017seqgan} uses GAN to generate sequences. Adversarial Variational Bayes \cite{mescheder2017adversarial} unifies GAN and VAE, which allows us to obtain a closer approximate posterior to the real posterior in VAE. VAEGAN \cite{2019VAEGAN} introduces AVB to train VAE and utilizes GAN for implicit variational inference, which provides a better approximation to the posterior and maximum likelihood assignment. To sum up, there is still a lack of research work in applying adversarial learning in sequential recommendation.

\begin{figure}[t]
    \centering
    \includegraphics[width=\linewidth]{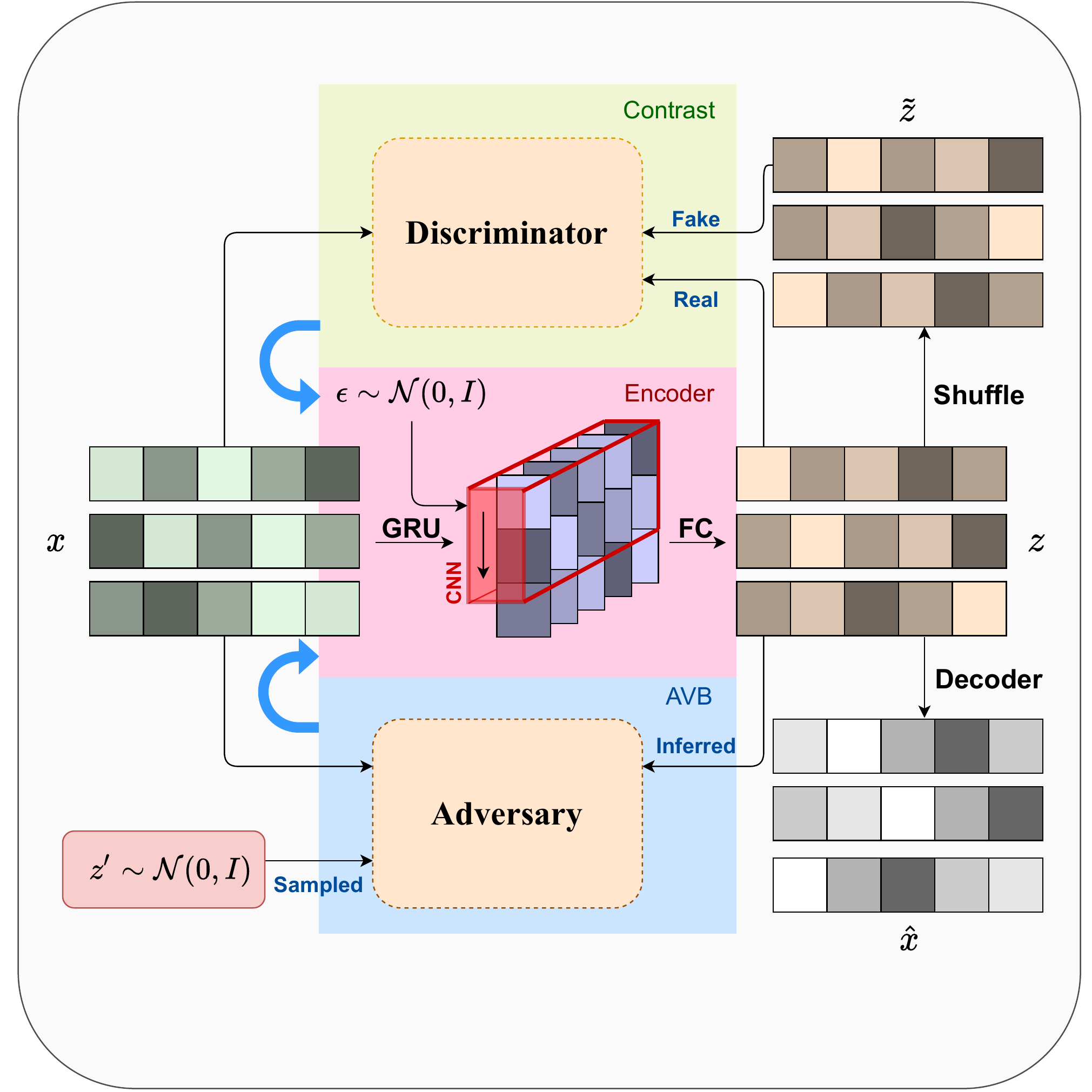}
    \caption{Structure of ACVAE. The model employs a CNN layer in the encoder of VAE. $(x,z)$ and $(x,\tilde{z})$ pairs are sent to discriminator, which calculates the contrastive loss. $(x,z)$ and $(x,z')$ pairs are sent to adversary, which calculates the approximate KL divergence. Both discriminator and adversary can optimize the parameters of the encoder through back-propagation.}
    \label{fig:model_struct}
\end{figure}

\section{The Method}
In this section, we first introduce the formulation of the sequential recommendation problem, then present our proposed method - Adversarial and Contrastive Autoencoder for Sequential Recommendation (ACVAE). Figure \ref{fig:model_struct} shows the structure of our proposed ACVAE. There are three main parts in ACVAE: contrast part, encoder part and adversary part. Original users' interaction sequence $\boldsymbol{x}$ is input to the encoder consists of RNN-CNN layers and noise function after embedding. The output of the encoder is $\boldsymbol{z}$, which is then input into the decoder, discriminator and adversary. The adversary receives inputs $(\boldsymbol{x},\boldsymbol{z})$ and $(\boldsymbol{x},\boldsymbol{z'})$, where $\boldsymbol{z'}$ is sampled from normal Gaussian distribution. The discriminator receives inputs $(\boldsymbol{x},\boldsymbol{z})$ and $(\boldsymbol{x},\tilde{\boldsymbol{z}})$, where $\tilde{\boldsymbol{z}}$ is the latent variable $\boldsymbol{z}$ after shuffled.

\subsection{Problem Formulation}
In this paper, $\mathcal{U}=\{u_1,u_2,\cdots,u_{|\mathcal{U}|}\}$ denotes the set of all users and $\mathcal{V} = \{v_1,v_2,\cdots,v_{|\mathcal{V}|}\} $ denotes the set of all items, where $|\mathcal{U}|$ and $|\mathcal{V}|$ denote the number of users and items respectively. $\boldsymbol{x}_{u}=\{\boldsymbol{x}_{u,1},\boldsymbol{x}_{u,2},\cdots,\boldsymbol{x}_{u,T_u}\}$ represents the interaction sequence of the user $u$, where $\boldsymbol{x}_{u,i}$ means the $i$-th item in the sequence that user $u$ interacts with and $T_u$ denotes the number of items in the user sequence. The sequential recommendation problem is to predict the next item (i.e. the $(T_u +1)$-th item) that user $u$ may be interested in given his historical interaction sequence $\boldsymbol{x}_u$.

\subsection{Sequential VAE Model}
VAE is a typical encoder-decoder structure and VAE has powerful generation capability. VAE models the latent variable $\boldsymbol{z}$ through the probability method, in which the latent variable $\boldsymbol{z}$ is a probability distribution and the observed data $\boldsymbol{x}$ can be reconstructed by a generative model (decoder) with the conditional probability $P_{\theta}(\boldsymbol{x}|\boldsymbol{z})$. In common VAE tasks, the target is to estimate the distribution of the latent variable $\boldsymbol{z}$ according to the observed data $\boldsymbol{x}$, which is called the inference of $\boldsymbol{z}$. Since it is usually difficult to calculate the posterior probability $P(\boldsymbol{z}|\boldsymbol{x})$, VAE employs an approximate posterior distribution $Q_{\phi}(\boldsymbol{z}|\boldsymbol{x})$ to fit the true posterior distribution of $\boldsymbol{z}$.

Mult-VAE \cite{liang2018variational} introduces VAE model into collaborative filtering. In Mult-VAE, the generative model generates the distribution of the all items $\pi(\boldsymbol{\hat{x}}_u)$ from a latent variable $\boldsymbol{z}_u$ sampled from a normal Gaussian distribution through a function $f_{\theta}$ which can be implemented with neural network. When making recommendations, $\boldsymbol{\hat{x}}_u$ is sampled from multinomial distribution with the distribution $\pi(\boldsymbol{\hat{x}}_u)$.

In the task of sequential recommendation, the modeling of the temporal dependencies between items is very important. The temporal dependencies can be modeled by a conditional probability. For a certain timestep $t$, the sequential model predicts the $(t+1)$-th item according to the items numbering from $1$ to $t$, and the conditional probability of the sequential model is $P(\boldsymbol{x}_{u,t+1}|\boldsymbol{x}_{u,[1:t]})$. Mult-VAE fails to model temporal order of items, but SVAE as a recursive version of VAE is proposed to capture the time dependence of the sequence. SVAE generates the target $\boldsymbol{x}_{u,t}$ for latent variable $\boldsymbol{z}_{u,t}$ at every timestep $t$:
\begin{align}
    \begin{split}
        \boldsymbol{z}_{u,t} & \sim\mathcal{N}(0,\boldsymbol{I}_{|\boldsymbol{z}_{u,t}|})\\
        \pi_{\theta}(\boldsymbol{z}_{u,t}) & \sim\exp(f_{\theta}(\boldsymbol{z}_{u,t}))\\
        \boldsymbol{x}_{u,t} & \sim Multi(1,\pi_{\theta}(\boldsymbol{z}_{u,t}))
    \end{split}
\end{align}
where $\mathcal{N}$ denotes the Gaussian distribution, $f_{\theta}$ denotes a function (usually a neural network) with parameter $\theta$ and $\pi_{\theta}$ is the distribution function of $f_{\theta}$ after softmax. The generated $x_{u,t}$ is sampled from multinomial distribution.

As for the inference model, we can obtain the approximate posterior distribution of latent variable $z_{u,t}$ according to the previous items $\boldsymbol{x}_{u,[1:t]}$ by the encoder. The approximate posterior distribution $Q_{\phi}(\boldsymbol{z}_{u,[1:T_u]}|\boldsymbol{x}_{u,[1:T_u]})$ can be factorized as:
\begin{equation}
    Q_{\phi}(\boldsymbol{z}_{u,[1:T_u]}|\boldsymbol{x}_{u,[1:T_u]})=\prod_{t=1}^{T_u}q_{\phi}(z_{u,t}|\boldsymbol{x}_{u,[1:t]})
\end{equation}

Then, the target sequence $\widehat{\boldsymbol{x}}_u$ can be generated by the decoder $P_{\theta}(\boldsymbol{x}_{u,[2:T_u+1]}|\boldsymbol{z}_{u,[1:T_u]})$:
\begin{align}
\begin{split}
    P(\hat{\boldsymbol{x}}_u,\boldsymbol{z}_u) & = P_{\theta}(\boldsymbol{x}_{u,[2:T_u+1]}|\boldsymbol{z}_{u,[1:T_u]})Q_{\phi}(\boldsymbol{z}_{u,[1:T_u]})\\
    & = \prod_{t=1}^{T_u}p_{\theta}(x_{u,t+1}|z_{u,t})q_{\phi}(z_{u,t})
\end{split}
\end{align}
in which the output $\hat{\boldsymbol{x}}_u$ is sampled from multinomial distribution.

\begin{figure}[t]
    \centering
    \subfigure[Real and Fake Distributions]{
    \begin{minipage}[c]{0.23\textwidth}
    \centering
    \includegraphics[width=\linewidth]{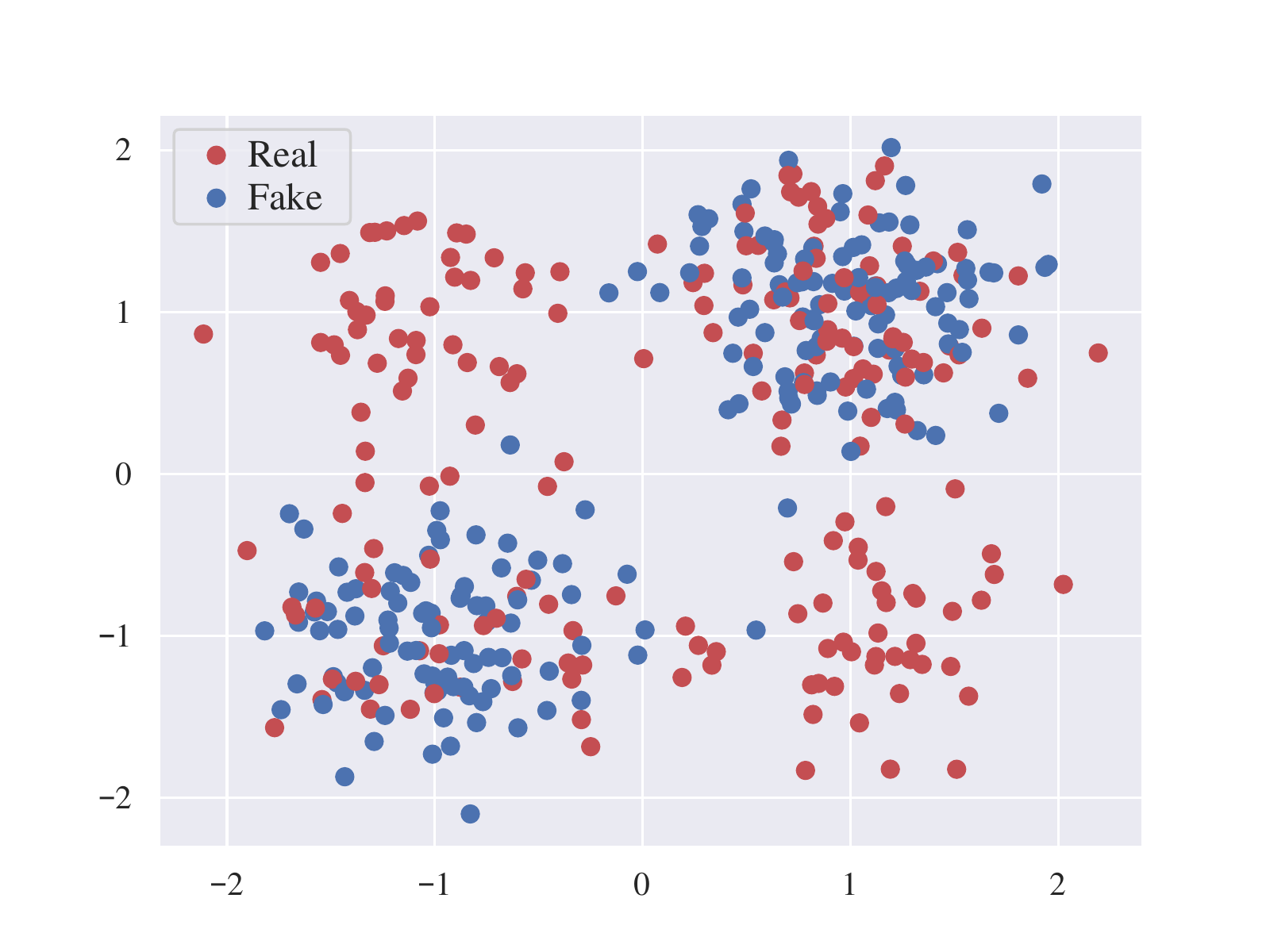}
    \label{fig:dependent}
    \end{minipage}%
    }
    \subfigure[Value of Discriminator]{
    \begin{minipage}[c]{0.23\textwidth}
    \centering
    \includegraphics[width=\linewidth]{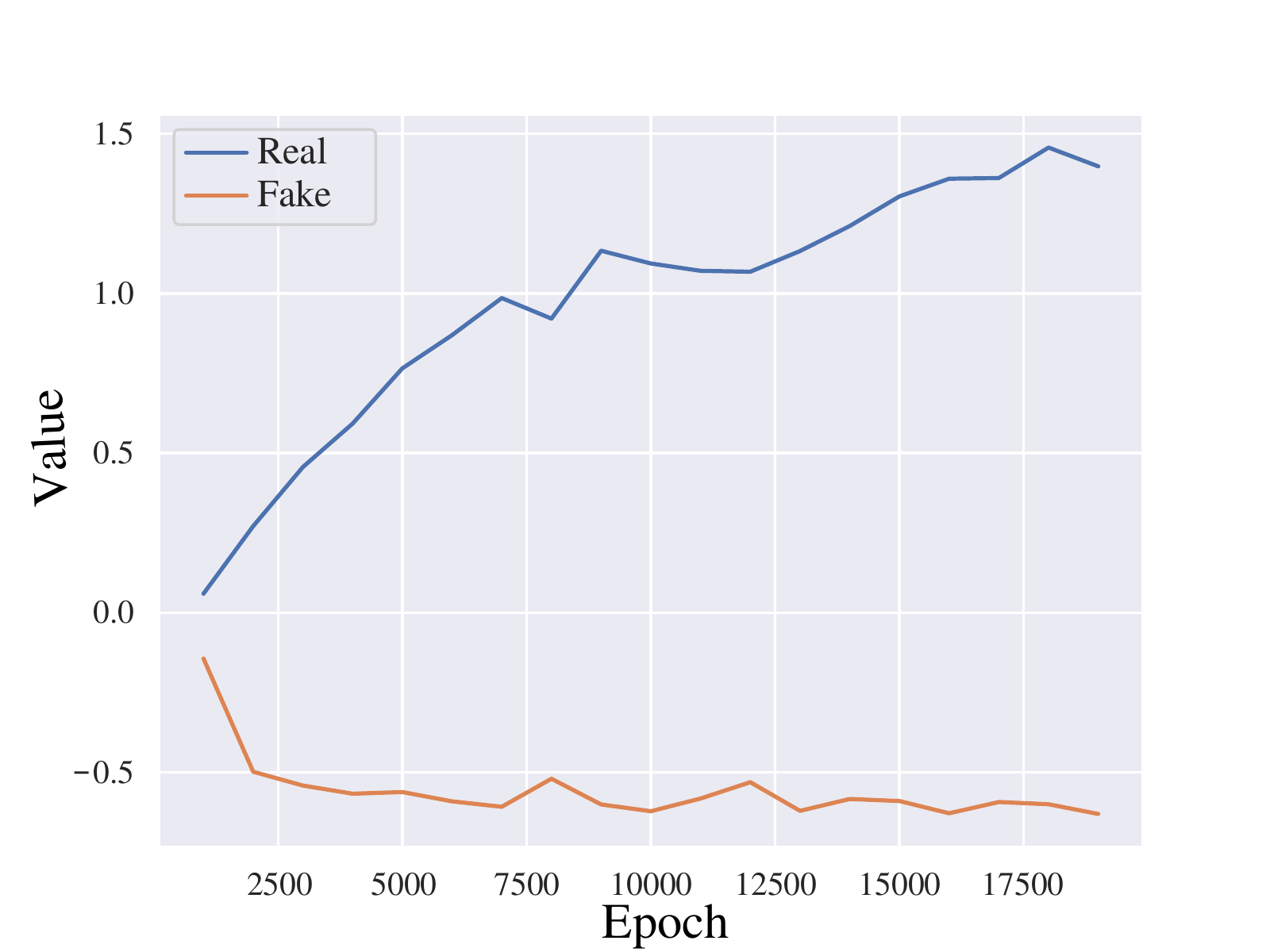}
    \label{fig:kl_avb}
    \end{minipage}%
    }
    \caption{Figure (a) shows the points sampled from distribution real (independent) and fake (dependent) distribution with $\sigma=0.4$. Figure (b) shows the output value of discriminator $D$ of the real and fake data.}
    \label{fig:vae_study}
\end{figure}

\subsection{Adversarial Learning for Sequential VAE}
In this section, we will first introduce the shortcomings of the traditional VAE models and show the important role of AVB in inferring latent variables with small correlation between different dimensions in VAE. Then, the AVB approach is proposed to improve the quality of latent variables for sequential recommendation. 

\subsubsection{Shortcomings of Traditional VAE Models}
The traditional VAE does sampling from the Gaussian distribution for the approximate posterior distribution $Q_{\phi}$ in the inference model, and thus leading to limited representational ability of latent variables. AVB has shown that the representational ability can be improved by inferring the latent variables in a flexible black box with adversarial training \cite{mescheder2017adversarial}. However, this is not the only difference between VAE and AVB. 

We use the following simple example to illustrate the problem. Consider a 2-dim latent variable $z$ sampled from $\mathcal{N}(\boldsymbol{\mu},{\boldsymbol{\sigma}^2})$, where $\mu$ and $\sigma$ are the outputs of a special neural network, which always outputs either $[-1,-1]$ or $[1,1]$ with equal probability for $\boldsymbol{\mu}$ and constant values for $\boldsymbol{\sigma}$. This is regarded as the fake data, which has strong correlation between the two dimensions. Then, consider the $\boldsymbol{\mu}$ of real data is the discrete uniform distribution of $\{-1,1\}\times\{-1,1\}$ and the real $z$ is also sampled from Gaussian distribution, and thus the two dimensions are independent. Since in VAE, the common approach for calculating KL divergence is simply calculating $\mathrm{KL}(Q(\boldsymbol{z}|\boldsymbol{x})||P(\boldsymbol{z}))$, where $P(\boldsymbol{z})$ is the prior distribution $\mathcal{N}(0,\boldsymbol{I})$ and $Q(\boldsymbol{z}|\boldsymbol{x})$ is the approximate posterior distribution. Each dimension of $\boldsymbol{z}$ is conditional independent. In this example, the KL divergence for the real and fake data are the same, since they share the same $\boldsymbol{\mu}$ and $\boldsymbol{\sigma}^2$. So the KL divergence can not distinguish the real and fake data (Figure \ref{fig:dependent}). However, if we input the real and fake data into an discriminator $D$ which is a simple fully connected neural network, it shows different answers for these two kinds of sampled points (Figure \ref{fig:kl_avb}).

The above example demonstrates that in some cases, discriminator may have certain advantages in judging the correlation of sampled dimensions compared with normal VAE. We will further show the influence of AVB on the correlation between different dimensions of latent variables in the experiment. This independence allows latent variables to capture more representative and disentangled characteristics of different users in recommendation. On the other hand, for sequential VAE models, encoder is supposed to capture the new information at each timestep to model the users' characteristics that change over time more accurately. The change of users' interest at different time may be subtle, which puts forward higher requirements for the expressiveness of the latent variables. AVB exploits the method of adversarial learning to increase the diversity of the distribution of latent variables and reducing the correlation of different dimensions between latent variables. That's why AVB is a suitable choice for our sequential VAE model.

\subsubsection{Adversarial Sequential Variational Bayes}
Because of the reasons above, we introduce adversarial learning into the sequential model and train the model in the framework of Adversarial Variational Bayes (AVB). AVB enables us to use complex inference models for VAE with adversarial learning, which generates diverse probability distributions that are close to the true posterior distribution. Sequential VAE performs maximum-likelihood training of variational evidence lower bound to estimate the intractable marginal log-likelihood  $\mathbb{E}_{\boldsymbol{x}_u\sim P_{\mathcal{D}}(\boldsymbol{x})} \log{P_{\theta}(\boldsymbol{x}_u)}$. For a single user $u$, we have:
\begin{align}
\begin{split}
    \log{P_{\theta}(\boldsymbol{\hat{x}}_u)} &\ge \mathbb{E}_{\boldsymbol{z}_u\sim Q_{\phi}(\boldsymbol{z}_u|\boldsymbol{x}_u)}\log{P_{\theta}}(\boldsymbol{\hat{x}}_u|\boldsymbol{z}_u)\\
    &\qquad\qquad\qquad\qquad\quad- \mathrm{KL}(Q_{\phi}(\boldsymbol{z}_u|\boldsymbol{x}_u)||P(\boldsymbol{z}_u))\\
    &=\sum_{t=1}^{T_u}[\mathbb{E}_{\boldsymbol{z}_{u,t}\sim q_{\phi}(\boldsymbol{z}_{u,t}|\boldsymbol{x}_{u,[1:t]})}\log{p_{\theta}}(\boldsymbol{x}_{u,t+1}|\boldsymbol{z}_{u,t})\\
    &\qquad\qquad\qquad-\mathrm{KL}(q_{\phi}(\boldsymbol{z}_{u,t}|\boldsymbol{x}_{u,[1:t]})||p(\boldsymbol{z}_{u,t}))]
    \label{eq:1}
\end{split}
\end{align}
where $P$ is the data distribution and $\theta, \phi$ stand for the parameters of generative and inference model respectively. The right side of the equation \eqref{eq:1} is called the evidence lower bound (ELBO).

Our goal is to optimize the marginal log-likelihood of $\boldsymbol{x}_u$. However, it is usually difficult to calculate directly because the parameter $\theta$ relies on $p_{\theta}(\boldsymbol{x}_u|\boldsymbol{z}_u)$. To solve this problem, we change the AVB term of objective function to:
\begin{align}
\begin{split}
    \max_{\theta,\phi}\sum_{t=1}^{T_u} \mathbb{E}_{\boldsymbol{x}_{u,t}\sim p_{\mathcal{D}}(\boldsymbol{x}_{u,t})} [\mathbb{E}&_{\boldsymbol{z}_{u,t} \sim q_{\phi}(\boldsymbol{z}_{u,t}|\boldsymbol{x}_{u,[1:t]})}  \log{p_{\theta}(\boldsymbol{x}_{u,t+1}|\boldsymbol{z}_{u,t})}\\
    & -\mathrm{KL}(q_{\phi}(\boldsymbol{z}_{u,t}|\boldsymbol{x}_{u,[1:t]})||p(\boldsymbol{z}_{u,t}))]
\end{split}
\label{eq:3}
\end{align}
Commonly, most of the VAE models set $q_{\phi}(\boldsymbol{z}_{u,t}|\boldsymbol{x}_{u,[1:t]})$ to be independent Gaussian distribution and parameterize it by means of the neural network. In this case, $\mathrm{KL}(q_{\phi}(\boldsymbol{z}_{u,t}|\boldsymbol{x}_{u,[1:t]})||p(\boldsymbol{z}_{u,t}))$ is easy to calculate because $q_{\phi}(\boldsymbol{z}_{u,t}|\boldsymbol{x}_{u,[1:t]})$ and $p(\boldsymbol{z}_{u,t})$ are Gaussian distribution and KL can be computed in an explicit way. But it will also make the model heavily rely on $\boldsymbol{z}_{u,t}$, which constraints the quality of the inference model. On the contrary, a higher quality of $q_{\phi}(\boldsymbol{z}_{u,t}|\boldsymbol{x}_{u,[1:t]})$ can result in a more flexible representational ability, as well as a better approximation to the true posterior \cite{huszar2017variational}. 
 
 As for the KL divergence, the approximate posterior distribution $q_{\phi}(\boldsymbol{z}_{u,t}|\boldsymbol{x}_{u,[1:t]})$ is intractable. However, the KL divergence can be restated as $\log q_{\phi}(\boldsymbol{z}_{u,t}|\boldsymbol{x}_{u,[1:t]})-\log p(\boldsymbol{z}_{u,t})$ and the AVB term of the objective function is given by:
\begin{align}
\begin{split}
    \max\limits_{\theta,\phi}\sum_{t=1}^{T_u} \mathbb{E}_{\boldsymbol{x}_{u,t}\sim p_{\mathcal{D}}(\boldsymbol{x}_{u,t})} &\mathbb{E}_{\boldsymbol{z}_{u,t} \sim  q_{\phi}(\boldsymbol{z}_{u,t}|\boldsymbol{x}_{u,[1:t]})}[ \log{p_{\theta}(\boldsymbol{x}_{u,t+1}|\boldsymbol{z}_{u,t})}\\
    &+\log p(\boldsymbol{z}_{u,t})-\log q_{\phi}(\boldsymbol{z}_{u,t}|\boldsymbol{x}_{u,[1:t]})]
    \label{eq:5}
\end{split}
\end{align}
To calculate equation \eqref{eq:5}, we first introduce a discriminative network $T_{\Psi}(\boldsymbol{x}_{u},\boldsymbol{z}_{u})$, which receives the whole sequence of $\boldsymbol{x}_u$ and $\boldsymbol{z}_u$ and outputs the sequence of values with length $T_u$. The discriminative network is originally used as a discriminator in Generative Adversarial Network (GAN) to check whether the generated item is real or not. Here we use it to distinguish the pair $(\boldsymbol{x}_{u},\boldsymbol{z}_{u})$ sampled with the posterior distribution $Q_{\phi}(\boldsymbol{z}_{u}|\boldsymbol{x}_{u})P_{\mathcal{D}}(\boldsymbol{x}_{u})$ from the pair sampled with the prior distribution $P(\boldsymbol{z}_{u})P_{\mathcal{D}}(\boldsymbol{x}_{u})$. As a result, we set the equation below as the discriminator $T_{\Psi}$ objective :

\begin{align}
\begin{split}
    &\max\limits_{\Psi}\sum_{t=1}^{T_u}[\mathbb{E}_{\boldsymbol{x}_{u}\sim P _{\mathcal{D}}(\boldsymbol{x}_{u})}\mathbb{E}_{\boldsymbol{z}_{u} \sim Q_{\phi}(\boldsymbol{z}_{u}|\boldsymbol{x}_{u})} \log (\sigma (T_{\Psi}(\boldsymbol{x}_{u},\boldsymbol{z}_{u})^{(t)}))\\
    &\qquad\qquad+ \mathbb{E}_{\boldsymbol{x}_{u}\sim P_{\mathcal{D}}(\boldsymbol{x}_{u})}\mathbb{E}_{\boldsymbol{z}_{u} \sim P(\boldsymbol{z}_u)} \log (1 -\sigma (T_{\Psi}(\boldsymbol{x}_{u},\boldsymbol{z}_{u})^{(t)}))]
    \label{eq:6}
\end{split}
\end{align}
where $\sigma (x) = (1+\mathrm{e}^{-x})^{-1}$ is the sigmoid function, and $\Psi$ denotes the parameters of the discriminative network $T_{\Psi}(\boldsymbol{x}_{u},\boldsymbol{z}_{u})$. When equation \eqref{eq:6} achieves its maximum, the optimal discriminative network $T^*(\boldsymbol{x}_{u},\boldsymbol{z}_{u})$ would be:
\begin{equation}
    T^*(\boldsymbol{x}_{u},\boldsymbol{z}_{u})^{(t)} = \log q_{\phi}(\boldsymbol{z}_{u,t}|\boldsymbol{x}_{u,t})-\log p(\boldsymbol{z}_{u,t}) 
    \label{eq:7}
\end{equation}
Substitute it into the previous AVB term of the objective function and the expression can be written as:
\begin{align}
\begin{split}
    \max_{\theta,\phi} \sum_{t = 1}^{T_u}\mathbb{E}_{\boldsymbol{x}_{u,t}\sim p_{\mathcal{D}}(\boldsymbol{x}_{u,t})} \mathbb{E}_{\boldsymbol{z}_{u,t}\sim q_{\phi}(\boldsymbol{z}_{u,t}|\boldsymbol{x}_{u,t})}[& \log{p_{\theta}(\boldsymbol{x}_{u,t+1}|\boldsymbol{z}_{u,t})}\\
    &\quad -T^*_{\Psi}(\boldsymbol{x}_{u},\boldsymbol{z}_{u})^{(t)}]
    \label{eq:8}
\end{split}
\end{align}
In practice, we usually consider the reparameterization trick \cite{kingma2013auto} so that parameters in this step can be optimized in the back propagation. Specifically, following \cite{2019VAEGAN}, we define a non-linear function $\boldsymbol{z}_{\phi}(\boldsymbol{x},\boldsymbol{\epsilon}) =f_1(f_{\epsilon}(f_2(\boldsymbol{x}),\boldsymbol{\epsilon}))$, where $f_1$ and $f_2$ denote the functions in the encoder and $f_{\epsilon}$ denotes the ``add $\epsilon$'' function (this part will be described in detail in Section 3.4). Both $\boldsymbol{\epsilon}$ are sampled from Gaussian distribution. Thus we can infer a flexible distribution with our proposed encoder by using the reparameterization trick. So finally, the AVB term of the objective function can be written as:
\begin{align}
\begin{split}
    \max\limits_{\theta,\phi} \sum_{t = 1}^{T_u}\mathbb{E}_{\boldsymbol{x}_{u,t}\sim p_{\mathcal{D}}(\boldsymbol{x}_{u,t})} \mathbb{E}_{\epsilon \sim \mathcal{N}(0,\mathcal{I})}[\log{P_{\theta}(\boldsymbol{x}_{u,t+1}|\boldsymbol{z}_{\phi}(\boldsymbol{x}_{u,[1:t]},\boldsymbol{\epsilon}))}\\ 
    -T^*_{\Psi}(\boldsymbol{x}_{u},\boldsymbol{z}_{\phi}(\boldsymbol{x}_{u},\boldsymbol{\epsilon}))^{(t)}]
\end{split}
\label{eq:9}
\end{align}

During training, we hope both of the objectives (both equation \eqref{eq:6} and \eqref{eq:9}) can get to their optimal values. However, they show the contrast tendency in optimizing any one of them. That is to say, one tends to get worse with the other closer to its optimal value. So we need alternate training in order to optimize both of them.

\subsection{Encoder Structure}
In order to bring sequential information for latent variables in ACVAE, we need a powerful encoder to capture potential temporal dependencies between items. In SVAE, a recurrent layer is employed to capture the dependencies between the current item and its previous ones. Although long-term sequence dependencies can be captured by RNN units like GRU or LSTM, local relationship between adjacent items may be ignored. In order to enhance the local relationship, we design a special convolutional layer combined with RNN as the encoder.
\subsubsection{RNN Layer}
When approximating posterior distribution $Q_{\phi}$, we use GRU as the RNN layer to capture long-distance dependencies. Thus we can get the latent variable $\boldsymbol{h}_{u,t}$ containing information about $\boldsymbol{x}_{u,[1:t]}$ at timestep $t$. This follows the design of SVAE. 
\begin{equation}
    \boldsymbol{h}_{u,t} = \xi(GRU_{\phi}(\boldsymbol{x}_{u,t},\boldsymbol{h}_{u,t-1}))
\end{equation}
where $\xi(x)=\log(1+\exp(x))$ is the softplus function.

\subsubsection{CNN Layer}
After the RNN layer, a convolutional layer is used to capture local features in the sequences in our model, where the collection of hidden states $\boldsymbol{h}_{u,t}$ in RNN will serve as the input of our convolutional layer. As shown in figure \ref{fig:model_struct}, a vertical CNN filter covers certain adjacent items in the user sequence, and convolves in the sequence to get local features.

Before convolution, the "add $\epsilon$" function $\boldsymbol{h}_{u,t}'=f_{\epsilon}(\boldsymbol{h}_{u,t},\boldsymbol{\epsilon})$ is applied to the output $\boldsymbol{h}_{u,t}$, which brings noise to the encoder of AVB. Then we define the filter $\boldsymbol{f}_v \in \mathbb{R}^{m \times 1}$. During convolution, the filter $\boldsymbol{f}_v$ moves on the plane of size $T_u\times d$ in both horizontal and vertical directions. In order to make the input $\boldsymbol{h}_{u,t}'$ and the output $\boldsymbol{c}_{u,t}$ correspond to each other on timestep $t$, we add several zero-padding rows. If the filter's bottom reached $(t+1)$-th row when it is generating the $t$-th output, the $(t+1)$-th item's information would be revealed to the $t$-th item, which results in label leakage. Therefore, we add all the zero-padding rows before our real data. As a result, we get the output matrix of the CNN layer $\boldsymbol{c}_{u,t}$.

\subsubsection{Fully Connected Layer}
Finally, we use a fully connected network to transform the output of the CNN layer to the latent variable $z$.
$$
    \boldsymbol{z}_{u,t}=\xi(\boldsymbol{c}_{u,t})\cdot\boldsymbol{W}+\boldsymbol{b}
$$
where $\boldsymbol{W}$ denotes the weight of the full connected layer and $\boldsymbol{b}$ denotes the bias.

\subsection{Contrastive Learning}
In sequential recommendation, the sequences of different users may be relatively similar, which makes it difficult for the model to analyze the user's unique personalized characteristics. The decoder is required to be able to reconstruct the input in original VAE model. However, this goal turns to be even more difficult \cite{deepInfomax, oord2018representation} when a whole sequence needs to be accurately reconstructed in SVAE. If the only goal of the model is to reconstruct the sequence, some truly effective and salient user's personalized information may be ignored. Here we to introduce contrastive learning to help train the sequential VAE model, which can also improve the "individuation" of latent variables. 

\subsubsection{Contrastive Loss}

We hope that the latent variables can obtain more effective and salient information about user $u$'s input $\boldsymbol{x}_u$ in our ACVAE model. In order to capture users' salient features, we learn the contrastive loss by employing a contrastive discriminator $G_{\omega}$ to compare the latent variables of different users.

Here, we define a pair $(\boldsymbol{x}_u, \boldsymbol{z}_u)$ is a positive match \emph{iff} $\boldsymbol{z}_u\sim Q_{\phi}(\boldsymbol{z}_u|\boldsymbol{x}_u)$, indicating $z_u$ is generated for user $x_u$. On the contrary, a pair $(\boldsymbol{x}_u, \widetilde{\boldsymbol{z}_u})$ is a negative match \emph{iff} $\widetilde{z_{u}}\sim Q_{\phi}(\boldsymbol{z}_{u'}|\boldsymbol{x}_{u'})$, where $u'\neq u$. The discriminator $G_{\omega}(\boldsymbol{x}_u, \boldsymbol{z}_u)$ aims to learn the relation between $\boldsymbol{x}_u$ and $\boldsymbol{z}_u$ and determine whether the pair $(\boldsymbol{x}_u, \boldsymbol{z}_u)$ comes from a positive match or a negative match.

In order to make $G_\omega$ to be able to distinguish the latent variables of different users, we need to find another term of the objective function which can measure the relationship between $\boldsymbol{x}$ and $\boldsymbol{z}$. Here, we define the contrastive loss $\mathcal{L}_{\omega, \phi}$ as follow:
\begin{align}
\begin{split}
    &\mathcal{L}_{\omega, \phi}(\boldsymbol{x}_u,\boldsymbol{z}_u)=\\
    &-\sum_{t = 1}^{T_u}[\mathbb{E}_{\boldsymbol{z}_{u}\sim Q_{\phi}(\boldsymbol{z}_{u}|\boldsymbol{x}_{u})}\log(\sigma(G_{\omega}(\boldsymbol{x}_{u},\boldsymbol{z}_{u})^{(t)}))\\
    &\qquad\quad+\mathbb{E}_{\widetilde{\boldsymbol{z}}_{u'}\sim Q_{\phi}(\tilde{\boldsymbol{z}_{u}}|\boldsymbol{x_{u'}})}\log(\sigma(1-G_{\omega}(\boldsymbol{x}_{u'},\widetilde{\boldsymbol{z}_{u}})^{(t)}))]
\end{split}
\end{align}
 When minimizing the contrastive loss $\mathcal{L}_{\omega,\phi}(\boldsymbol{x}_u,\boldsymbol{z}_u)$, the discriminator $G_\omega$ can better distinguish the positive matches and negative matches. The latent variables inferred by the encoder will obtain more salient and personalized information of different users.

\subsubsection{Optimization}
The optimization goal of contrastive learning part is minimizing the contrastive loss term of the objective function by optimizing $\omega$ and $\phi$. Since the encoder and the discriminator $G_{\omega}$ are both optimizing the contrastive term of the objective function, we do not need additional alternate training process. Thus this term can be simply added to the original term of the objective function in variational autoencoder when optimizing the parameters in the encoder.

\subsection{Objective Functions}
Overall, the whole objective functions of our proposed ACVAE for user $u$ are as follows:
\begin{align}
\begin{split}
    &\max_{\phi,\theta,\omega}\sum_{t = 1}^{T_u}\mathbb{E}_{\boldsymbol{x}_{u,t}\sim p_{\mathcal{D}}(\boldsymbol{x}_{u,t})} \mathbb{E}_{\boldsymbol{\epsilon}\sim\mathcal{N}(0,\mathcal{I})}[ \log{P_{\theta}(\boldsymbol{x}_{u,t+1}|\boldsymbol{z}_{\phi}(\boldsymbol{x}_{u,[1:t]},\boldsymbol{\epsilon}))}\\
    &\qquad\qquad-\alpha\cdot T^*_{\Psi}(\boldsymbol{x}_{u},\boldsymbol{z}_{\phi}(\boldsymbol{x}_{u},\boldsymbol{\epsilon}))^{(t)}-\beta\cdot\mathcal{L}_{\omega, \phi}(\boldsymbol{x}_{u},\boldsymbol{z}_{\phi}(\boldsymbol{x}_u,\epsilon))^{(t)}],\\
    &\max\limits_{\Psi}\sum_{t = 1}^{T_u}[\mathbb{E}_{\boldsymbol{x}_{u}\sim P _{\mathcal{D}}(\boldsymbol{x}_{u})}\mathbb{E}_{\boldsymbol{\epsilon}\sim\mathcal{N}(0,\mathcal{I})}\log (\sigma (T_{\Psi}(\boldsymbol{x}_{u},\boldsymbol{z}_{\phi}(\boldsymbol{x}_u,\boldsymbol{\epsilon}))^{(t)}))\\
    &\qquad\qquad\quad\,\,\,+\mathbb{E}_{\boldsymbol{x}_{u}\sim P _{\mathcal{D}}(\boldsymbol{x}_{u})}\mathbb{E}_{\boldsymbol{z}_{u}\sim P(\boldsymbol{z}_u)}\log(1-\sigma (T_{\Psi}(\boldsymbol{x}_{u},\boldsymbol{z}_u)^{(t)}))]
\end{split}
\end{align}
where $\alpha$ and $\beta$ are hyper-parameters controlling the weight of the discriminative and contrastive terms respectively. The pseudo code of our algorithm is shown in Algorithm \ref{alg1}, where $\eta$ is a hyper-parameter representing the learning rate.

\begin{algorithm}
    \caption{Training of ACVAE}
    \label{alg1}
    $i \gets 0$\;
    \While{$i<\mathrm{MaxIteration}$}{
        Fetch $u = [u_1, u_2,...,u_m]^T \subset \mathcal{U}$ \;
        Sample $\epsilon = [\epsilon_1,\epsilon_2,...,\epsilon_m]^T, \epsilon _{i} \sim \mathcal{N}(0,\boldsymbol{I})$ \;
        \eIf{$i \% 2 =0$}{
            // Compute $\phi, \theta, \omega$ - gradient:\\
                $g_{\phi}, g_{\theta}, g_{\omega} \gets  \frac{1}{m} \sum\limits_{k=1}^{m}\sum\limits_{t=1}^{T_{u_k}}\nabla_{\phi,\theta, \omega}[\log{P_{\theta}(\boldsymbol{x}_{u_k}|\boldsymbol{z}_{\phi}(\boldsymbol{x}_{u_k},\boldsymbol{\epsilon_k})^{(t)})}$\\
                $\qquad\qquad\qquad\qquad-\alpha\cdot T^*_{\Psi}(\boldsymbol{x}_{u_k},\boldsymbol{z}_{\phi}(\boldsymbol{x}_{u_k},\boldsymbol{\epsilon_k})^{(t)}$\\
                $\qquad\qquad\qquad\qquad-\beta\cdot\mathcal{L}_{\omega,\phi}(\boldsymbol{x}_{u_k},\boldsymbol{z}_{\phi}(\boldsymbol{x}_{u_k},\boldsymbol{\epsilon_k})^{(t)})]$\;
            // Perform SGD-updates for $\phi, \theta, \omega$ :\\
                $\phi \gets \phi + \eta g_{\phi},\,\theta \gets \theta +\eta g_{\theta},\,\omega \gets \omega + \eta g_{\omega}$\;
        }{
            // Compute $\Psi$ - gradient: \\
                $g_{\Psi} \gets \frac{1}{m} \sum\limits_{k=1}^{m}\sum\limits_{t=1}^{T_{u_k}} \nabla_{\Psi}[\log (\sigma(T_{\Psi}(\boldsymbol{x}_{u_k},\boldsymbol{z}_{\phi}(\boldsymbol{x}_{u_k},\boldsymbol{\epsilon_k})^{(t)})))$\\
                $\qquad\qquad\qquad\qquad+\log (1 -\sigma (T_{\Psi}(\boldsymbol{x}_{u_k},\boldsymbol{z}_{u_k})^{(t)}))]$\;
                \BlankLine
            // Perform SGD-updates for $\Psi$:\\
                $\Psi \gets \Psi + \eta g_{\Psi}$\\
        }
        $i\gets i + 1$\;
    }

\end{algorithm}

\section{Experiments}
\subsection{Datasets}
We evaluate our model on the following datasets, which are widely used in evaluating the performance of recommender system.

\begin{itemize}
    \item \textbf{MovieLens Latest (ML-latest)\footnote{https://grouplens.org/datasets/movielens}:} MovieLens Latest is a widely used dataset which contains the latest movie ratings with detailed timestamps. Ratings range from 1 to 5.
    \item \textbf{MovieLens 1m (ML-1m)\footnote{https://grouplens.org/datasets/movielens}:} This is a widely used dataset which contains 1 million ratings with detailed timestamps and rating level from 1 to 5.
    \item \textbf{MovieLens 10m (ML-10m)\footnote{https://grouplens.org/datasets/movielens}:} This is a larger version of ML-1m, which contains 10 million movie ratings by the users.
    \item \textbf{Yelp\footnote{https://www.yelp.com/dataset}:} Yelp contains businesses, reviews and user data including ratings and timestamps. We use a subset of the review data of it, which contains detailed review information since 2018.
\end{itemize}

\begin{table}[t]
	\centering
	\caption{Statistics of Preprocessed Datasets}
	\label{tab:data}
	\begin{tabular}{cccccc}
		\toprule
		Datasets&Records&Users&Items&Avg. Length&Sparsity \\
		\midrule
		ML-latest&46K&604&7.4K&76.2&98.9\%\\
		ML-1m&580K&6K&3.5K&95.3&97.3\%\\
		ML-10m&4.7M&69K&10K&67.8&99.3\%\\
	    Yelp&557K&51K&32K&11.0&99.9\%\\
		\bottomrule
	\end{tabular}
\end{table}

All the datasets are preprocessed following the similar approach in SVAE \cite{SVAE}. First, we only consider the interactions with rating score strictly larger than 3 as positive interactions on the datasets with rating values range from 1 to 5. Then, we remove the users interacted less than 5 times, as well as the items interacted less than 5 times. Since in our model, we only need the information of implicit feedback, the detailed numeric rating numbers are all set to 1. The unused item labels are ignored and all of the ratings or purchases are treated as interactions. Table \ref{tab:data} shows the statistics of our preprocessed datasets. We can see that these datasets' sizes and the average length of sequences differ significantly. The ML-10m dataset contains the most records while the ML-latest dataset contains the least records. And the average length of ML-latest, ML-1m and ML-10m are far more than that of Yelp. It enables us to further explore each model's performances on datasets of different sizes and different average lengths of sequences. 

\renewcommand\arraystretch{0.8}
\begin{table*}[!h]
\caption{Comparison between our proposed model and other baselines at top-$k$ when $k\in\{5,10,20\}$. Boldface denotes the highest scores and underlines denote the highest scores other than the best scores.}
\label{tab:1}
\centering
\begin{tabular}{c|ccc|ccc|ccc}
\toprule
\textbf{ML-latest} & Recall@5 & NDCG@5 & MRR@5 & Recall@10 & NDCG@10 & MRR@10 & Recall@20 & NDCG@20 & MRR@20 \\ \midrule
POP &0.028  &0.057  &0.114  &0.045  &0.055  &0.123  &0.070  & 0.062 &0.130  \\
FPMC & \underline{0.044} & 0.068 & 0.132 & \underline{0.077} & \underline{0.077} & \underline{0.151} & \underline{0.120} &\underline{0.091} & \underline{0.161}  \\
Caser & 0.027 & 0.049  &0.097  &0.058  &0.060  &0.116  &0.107  &0.077  &0.129  \\
GRU4Rec$^+$ &0.039  & \underline{0.071}  & \underline{0.136}  &0.065  &0.075  & \underline{0.151}  &0.105  &0.086  &0.160  \\
BERT4Rec &0.038  &0.056  &0.110  &0.059  &0.057  &0.122  &0.087  &0.061  &0.129 \\
CFGAN &0.030  &0.065  &0.123  &0.060  &0.070  &0.139  &0.098  &0.080  &0.149  \\
Mult-VAE &0.033  &0.063  &0.128  &0.060  &0.066  &0.145  &0.111  &0.080  &0.154  \\
SVAE & 0.043 & 0.065  & 0.123 & 0.075 & 0.074  & 0.139 & \underline{0.120} & 0.089 & 0.148 \\
\textbf{ACVAE} & \textbf{0.052} & \textbf{0.081}  & \textbf{0.155}  & \textbf{0.091} & \textbf{0.092}  & \textbf{0.170}  & \textbf{0.145} & \textbf{0.107}  & \textbf{0.179}  \\\bottomrule
\end{tabular}

\begin{tabular}{c|ccc|ccc|ccc}
\toprule
\textbf{ML-1m} & Recall@5 & NDCG@5 & MRR@5 & Recall@10 & NDCG@10 & MRR@10 & Recall@20 & NDCG@20 & MRR@20 \\ \midrule
POP &0.024  &0.077  &0.141  &0.043  &0.076  &0.155  &0.088  & 0.087 &0.168  \\
FPMC & 0.065 & 0.133 & 0.244 & 0.113 & 0.141 & 0.267 & 0.180 & 0.158 & 0.277  \\
Caser & 0.066 & 0.139  &0.248  &0.114  &0.146  &0.271  &0.186  &0.164  &0.281  \\
GRU4Rec$^+$ &0.064  &0.125  &0.228  &0.104  &0.127  &0.247  &0.165  &0.141  &0.257  \\
BERT4Rec &0.053  &0.104  &0.196  &0.092  &0.110  &0.218  &0.154  &0.128  &0.230  \\
CFGAN &0.033  &0.073  &0.140  &0.062  &0.079  &0.160  &0.111  &0.094  &0.172  \\
Mult-VAE &0.029  &0.060  &0.116  &0.063  &0.069  &0.139  &0.114  &0.085  &0.152  \\
SVAE & \underline{0.077} &\underline{0.152}  &\underline{0.267}  & \underline{0.129} &\underline{0.157}  &\underline{0.289}  & \underline{0.207} &\underline{0.176}  &\underline{0.299}  \\
\textbf{ACVAE} & \textbf{0.095} & \textbf{0.188}  & \textbf{0.320}  & \textbf{0.158} & \textbf{0.192}  & \textbf{0.339}  & \textbf{0.244} & \textbf{0.212}  & \textbf{0.348}  \\\bottomrule
\end{tabular}

\begin{tabular}{c|ccc|ccc|ccc}
\toprule
\textbf{ML-10m} & Recall@5 & NDCG@5 & MRR@5 & Recall@10 & NDCG@10 & MRR@10 & Recall@20 & NDCG@20 & MRR@20 \\ \midrule
POP &0.032&0.056& 0.102&0.055 & 0.060&0.114  &0.088 &0.069  & 0.122 \\
FPMC & 0.058 & 0.090 & 0.163 & 0.100 & 0.101 & 0.183 & 0.169 & 0.123 & 0.195 \\
Caser &0.037&0.065&0.125 &0.067 & 0.072&0.142  &0.118 &0.088  & 0.153 \\
GRU4Rec$^+$ &0.080&0.122&0.214 &0.130 &0.132 &0.232  &0.199 &0.152  &0.241  \\
BERT4Rec &0.084&0.129&0.223 &0.143 &0.144 &0.244  &0.228 & 0.169 &0.255  \\
CFGAN &0.039&0.069&0.128 &0.067 &0.074 &0.143  &0.107 &0.084  &0.153  \\
Mult-VAE &0.049 &0.082&0.152 &0.092 &0.094 &0.173  &0.166 & 0.117& 0.186 \\
SVAE & \underline{0.109} & \underline{0.173} & \underline{0.295} & \underline{0.172} &\underline{0.184} & \underline{0.316} & \underline{0.260} & \underline{0.208} & \underline{0.325} \\
\textbf{ACVAE}& \textbf{0.112} & \textbf{0.177} & \textbf{0.302} & \textbf{0.180} & \textbf{0.190} & \textbf{0.321} & \textbf{0.270} & \textbf{0.215} & \textbf{0.330} \\\bottomrule
\end{tabular}

\begin{tabular}{c|ccc|ccc|ccc}
\toprule
\textbf{Yelp} & Recall@5 & NDCG@5 & MRR@5 & Recall@10 & NDCG@10 & MRR@10 & Recall@20 & NDCG@20 & MRR@20 \\ \midrule
POP &0.005  &0.004  &0.004  &0.009  &0.005  &0.005  &0.016  &0.007  &0.006  \\
FPMC & 0.011 & 0.007 & 0.008  & 0.019 & 0.010 & 0.011 & 0.034 & 0.015 & 0.012 \\
Caser &0.008  &0.006  &0.007  &0.014  &0.008  &0.008  &0.023  &0.010  &0.011  \\
GRU4Rec$^+$ &0.011  &0.008  &0.009 &0.020  & 0.011 &0.011  &0.034  &0.015  &0.013  \\
BERT4Rec &0.019  &0.013  & \underline{0.017}  &0.034  &0.019  &0.020  &0.056  &0.025  &0.022  \\
CFGAN &0.004  &0.003  &0.003  &0.007  &0.004  &0.004  &0.010  &0.005  &0.004  \\
Mult-VAE & \underline{0.022}  & \underline{0.015}  & 0.016  & 0.034  &0.019  &0.019 &0.056  &0.025  &0.021  \\
SVAE & 0.021 & \underline{0.015}  & \textbf{0.018}  & \underline{0.035} & \underline{0.020}  & \underline{0.021}  & \underline{0.057} & \underline{0.026}  & \underline{0.023}  \\
\textbf{ACVAE} & \textbf{0.023} & \textbf{0.016}  & \textbf{0.018}  & \textbf{0.039} & \textbf{0.021}  & \textbf{0.022}  & \textbf{0.066} & \textbf{0.028}  & \textbf{0.025}  \\\bottomrule
\end{tabular}
\end{table*}

\subsection{Baselines}
In order to evaluate the performance of our model, we take some models for recommendation as baselines, including the traditional approach according to popularity, RNN-based methods, attention-based method, adversarial learning method and VAE methods. This is a brief introduction of these methods:
\begin{itemize}
    \item \textbf{POP:} POP is a simple recommendation algorithm. It sorts by the number of user interactions and always recommends the ones with highest popularity.
    \item \textbf{FPMC} \cite{FPMC}: FPMC combines matrix factorization and markov chains together to recommend the items.
    \item \textbf{Caser} \cite{Caser}: Caser uses horizontal and vertical CNN to capture the information.
     \item \textbf{GRU4Rec$^+$} \cite{hidasi2015sessionbased}: GRU4Rec$^+$ uses GRU and new loss functions for session based recommendation.
    \item \textbf{BERT4Rec} \cite{BERT4Rec}: BERT4Rec employs BERT model, which trains the model by predicting the masked items with a bidirectional self-attention network.
    \item \textbf{CFGAN} \cite{CFGAN}: CFGAN uses a GAN structure to generate user's purchase vector.
    \item \textbf{Mult-VAE} \cite{liang2018variational}: Mult-VAE uses a multinomial likelihood for variational autoencoder to improve the performance.
    \item \textbf{SVAE} \cite{SVAE}: Sequential variational autoencoder uses GRU and variational autoencoder to generate the target sequence.

\end{itemize}

\subsection{Training Details}
We implement ACVAE with \texttt{PyTorch}. For POP, FPMC, GRU4Rec$^+$, CFGAN and Caser, we use the code in the NeuRec\footnote{https://github.com/wubinzzu/NeuRec} algorithm library. For SVAE, we use the code provided by the author. For Mult-VAE and BERT4Rec, we implement them with \texttt{PyTorch} following the original code.

The ACVAE model includes an embedding layer of size 128, a GRU layer of size 100, and the latent variables of size 64. A residual structure is added to the the convolutional layer in the encoder to prevent degradation of the gradient. Since the length of the item sequences of different users are not the same, we set several fixed sequence lengths $M$ for each datasets to gather different sequences in one batch. If the length of the sequence is larger than $M$, then we only keep the last $M$ items. If the length is smaller than $M$, we pad zeros to the end of item sequence. In our experiments, we use Adam optimizer with $\eta=1.0\times 10^{-4}$ and weight decay $l_2=1.0\times10^{-2}$ for VAE and SGD optimizer with $\eta=5.0\times10^{-4}$ and $l_2=1.0\times10^{-1}$ for $T_{\Psi}$ and $G_{\omega}$. The settings of $\alpha$ and $\beta$ have an impact on the experimental results, which will be further explored in Section 4.7.

For the sake of fairness, we set the embedding size to 128 for the models with neural network in the baselines except FPMC (embedding size equals to 64) for its poor performance when embedding size equals to 128. To make the model converge, the models are trained for 300 epochs on ML-latest, 200 epochs on ML-1m, 100 epochs on ML-10m and 40 epochs on Yelp. The source code of ACVAE is available on GitHub\footnote{https://github.com/ACVAE/ACVAE-PyTorch}.

\subsection{Evaluation Metrics}
The goal of our proposed method is predicting the next item that will be picked by the user. For each user, we split the interact sequence by 8:2 for training and testing respectively. After training, the users' training sets are used as test input and the output of last item will be taken as prediction.

We use some of the widely used metrics to evaluate the performance of our model for recommendation. We evaluate the metrics for top-$k$ item recommendation.
\begin{itemize}
    \item \emph{Recall:} Recall ratio of the ground truth item.
    $$
    Recall@k=\frac{Hits@k}{|L|}
    $$
    where $L$ denotes the relevant items in the test dataset.
    \item \emph{NDCG:} Normalized Discounted Cumulative Gain. $NDCG$ considers not only how many hits are recommended, but also the position the hits locates in top-k recommendation. The more the hits are and the closer the hits are to the top, the higher the score of $NDCG$ will be. 
    $$
    NDCG@k=\frac{DCG@k}{IDCG@k}=\frac{\sum_{i=1}^k\frac{1}{\log_2(i+1)}}{\sum_{i=1}^{|L|}\frac{1}{\log_2(i+1)}}    
    $$
    where $DCG = \sum_{i=1}^k\frac{1}{\log_2(i+1)}$ means Discounted Cumulative Gain and $IDCG = \sum_{i=1}^{|L|}\frac{1}{\log_2(i+1)}$ means the maximum of $DCG$.
    \item \emph{MRR:} Mean Reciprocal Rank of the ground truth item in the sorted prediction sequence.
    $$
    MRR@k=\frac{1}{r_f}
    $$
    where $r_f$ denotes the rank of the first hit in the prediction list.
\end{itemize}

\subsection{Performance Comparison}
We compare our ACVAE with those baselines, table \ref{tab:1} shows
the top-$k$ recommendation performance on the four datasets, where $k\in\{5, 10, 20\}$. The result shows that our proposed ACVAE can outperform other methods over all of the evaluation metrics. 

Compared with the VAE models (\emph{i.e.}, SVAE and Mult-VAE), our model has a significant improvement. That's because our model employs AVB and contrastive loss, which brings significant improvement to the inference of latent variables in sequential recommendation. 

\begin{figure*}[!htbp]
\centering
\subfigure[ML-latest]{
\begin{minipage}[c]{0.24\textwidth}
\centering
\includegraphics[width=\linewidth]{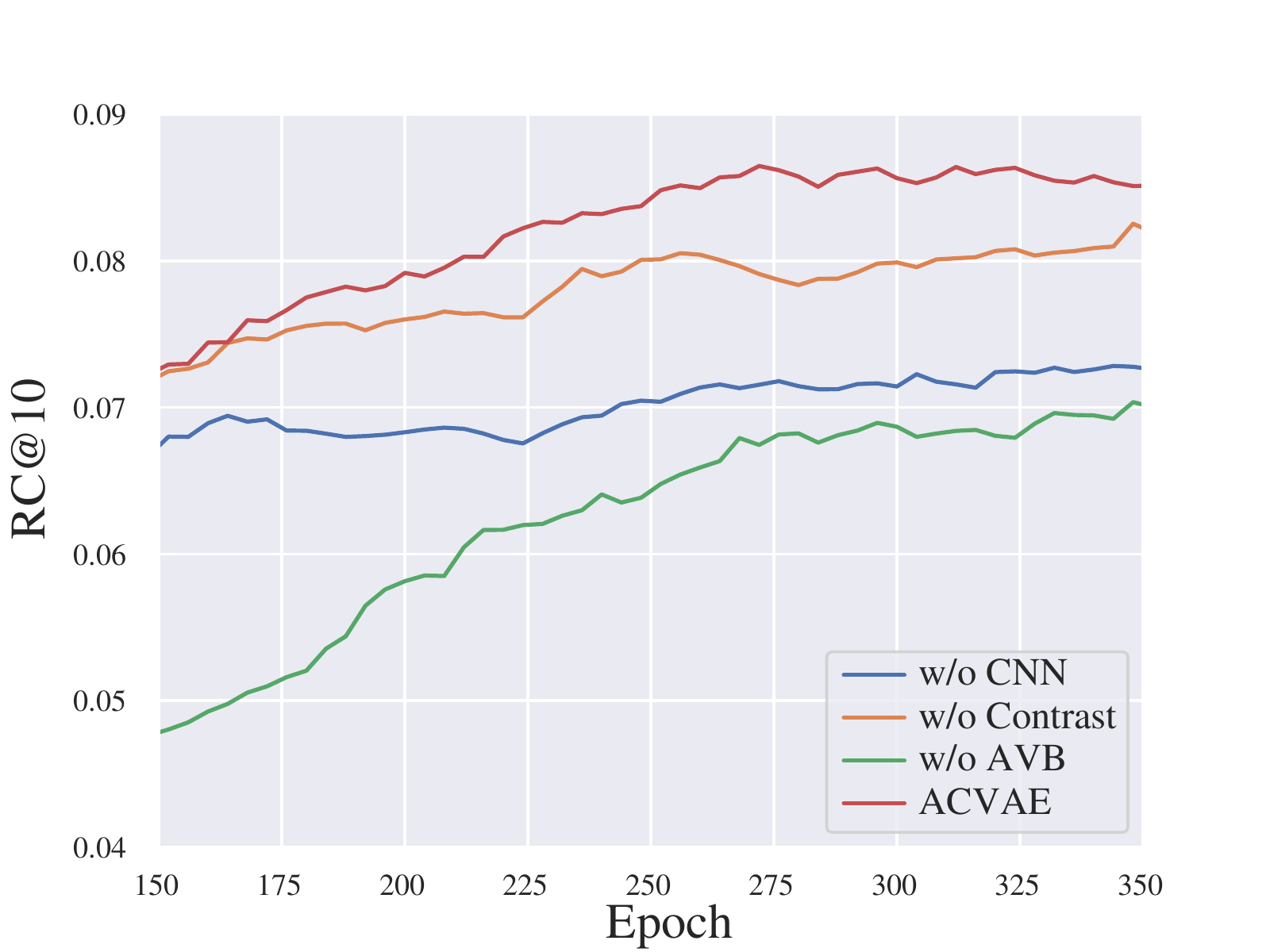}
\label{fig:ml-latest}
\end{minipage}%
}
\subfigure[ML-1m]{
\begin{minipage}[c]{0.24\textwidth}
\centering
\includegraphics[width=\linewidth]{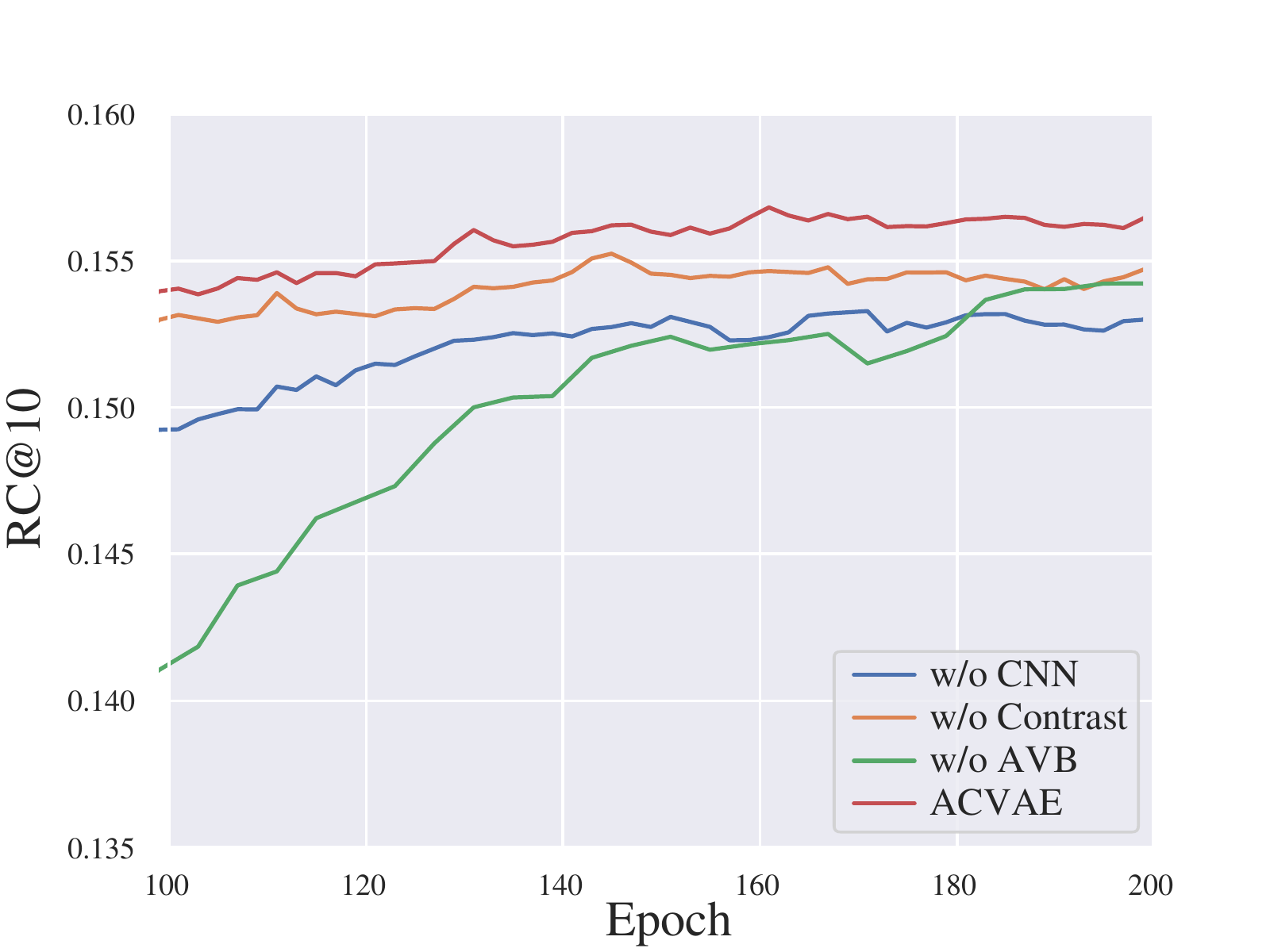}
\label{fig:ml-1m}
\end{minipage}%
}\subfigure[ML-10m]{
\begin{minipage}[c]{0.24\textwidth}
\centering
\includegraphics[width=\linewidth]{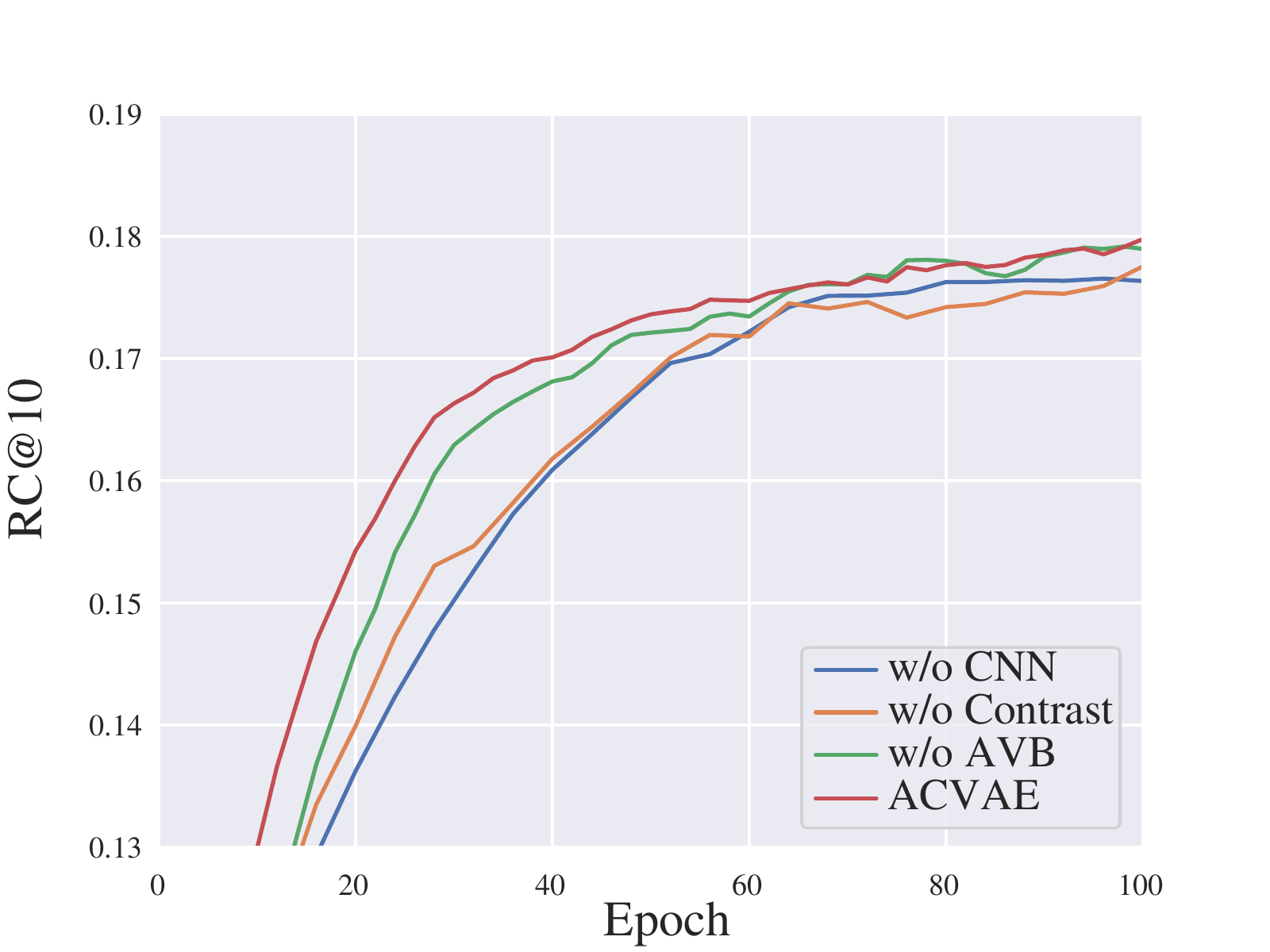}
\label{fig:ml-10m}
\end{minipage}%
}\subfigure[Yelp]{
\begin{minipage}[c]{0.24\textwidth}
\centering
\includegraphics[width=\linewidth]{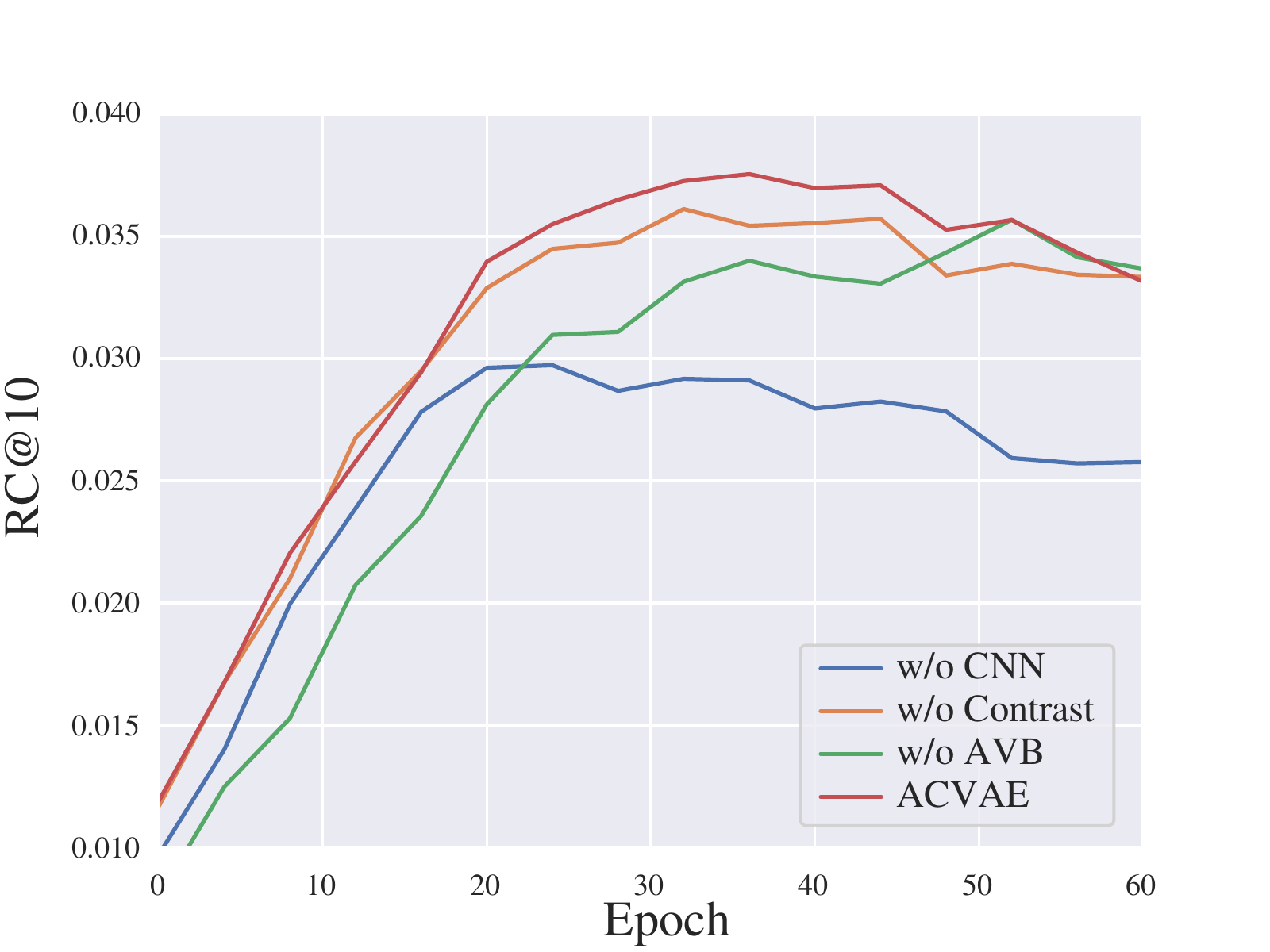}
\label{fig:yelp}
\end{minipage}%
}
\caption{Ablation study on ML-latest, ML-1m, ML-10m and Yelp.}
\label{fig:ablation-study}
\end{figure*}

\begin{figure*}[!htbp]
\centering
\subfigure[ML-latest]{
\begin{minipage}[c]{0.24\textwidth}
\centering
\includegraphics[width=\linewidth]{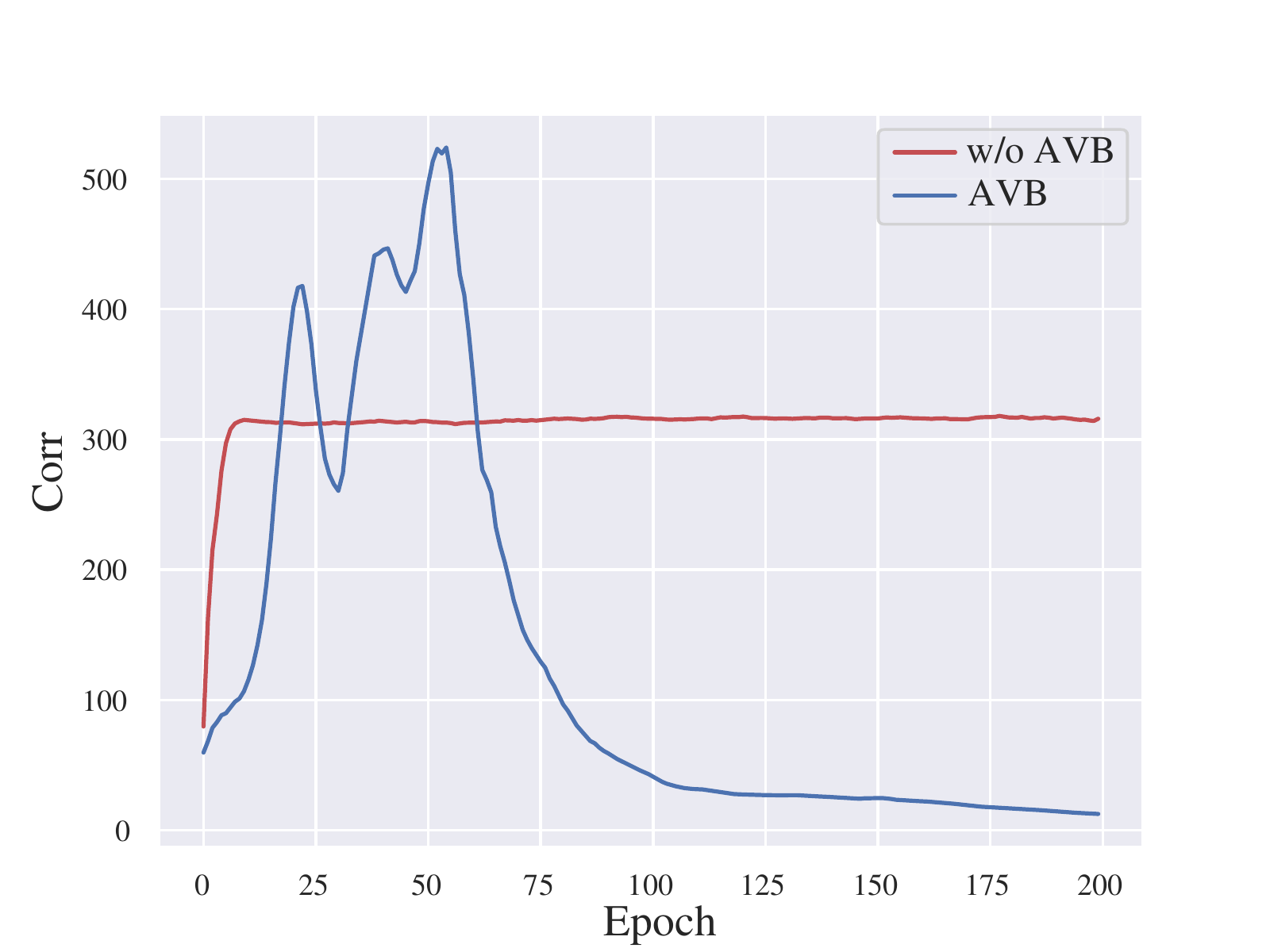}
\label{fig:corr-latest}
\end{minipage}%
}
\subfigure[ML-1m]{
\begin{minipage}[c]{0.24\textwidth}
\centering
\includegraphics[width=\linewidth]{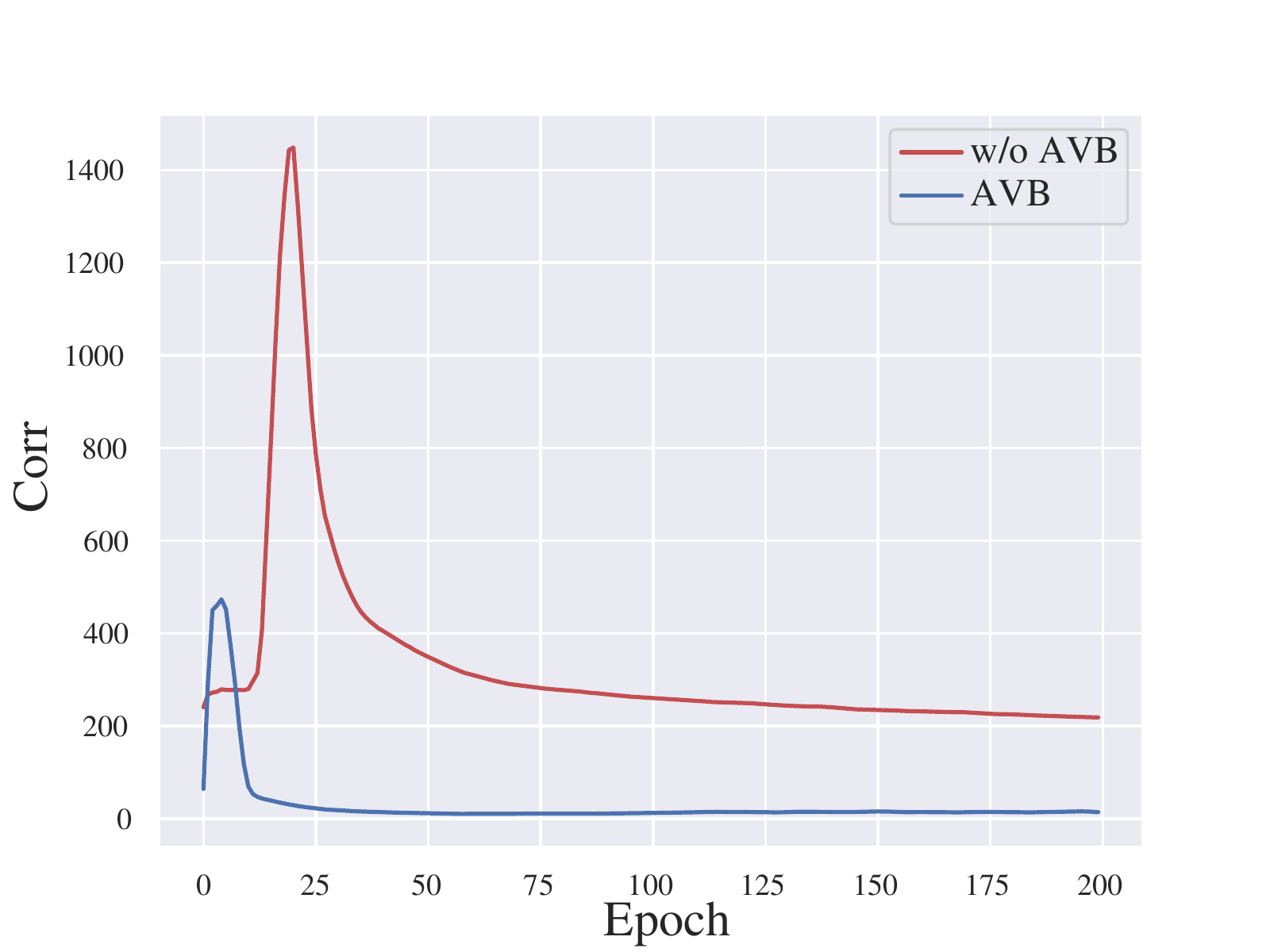}
\label{fig:corr-1m}
\end{minipage}%
}
\subfigure[ML-10m]{
\begin{minipage}[c]{0.24\textwidth}
\centering
\includegraphics[width=\linewidth]{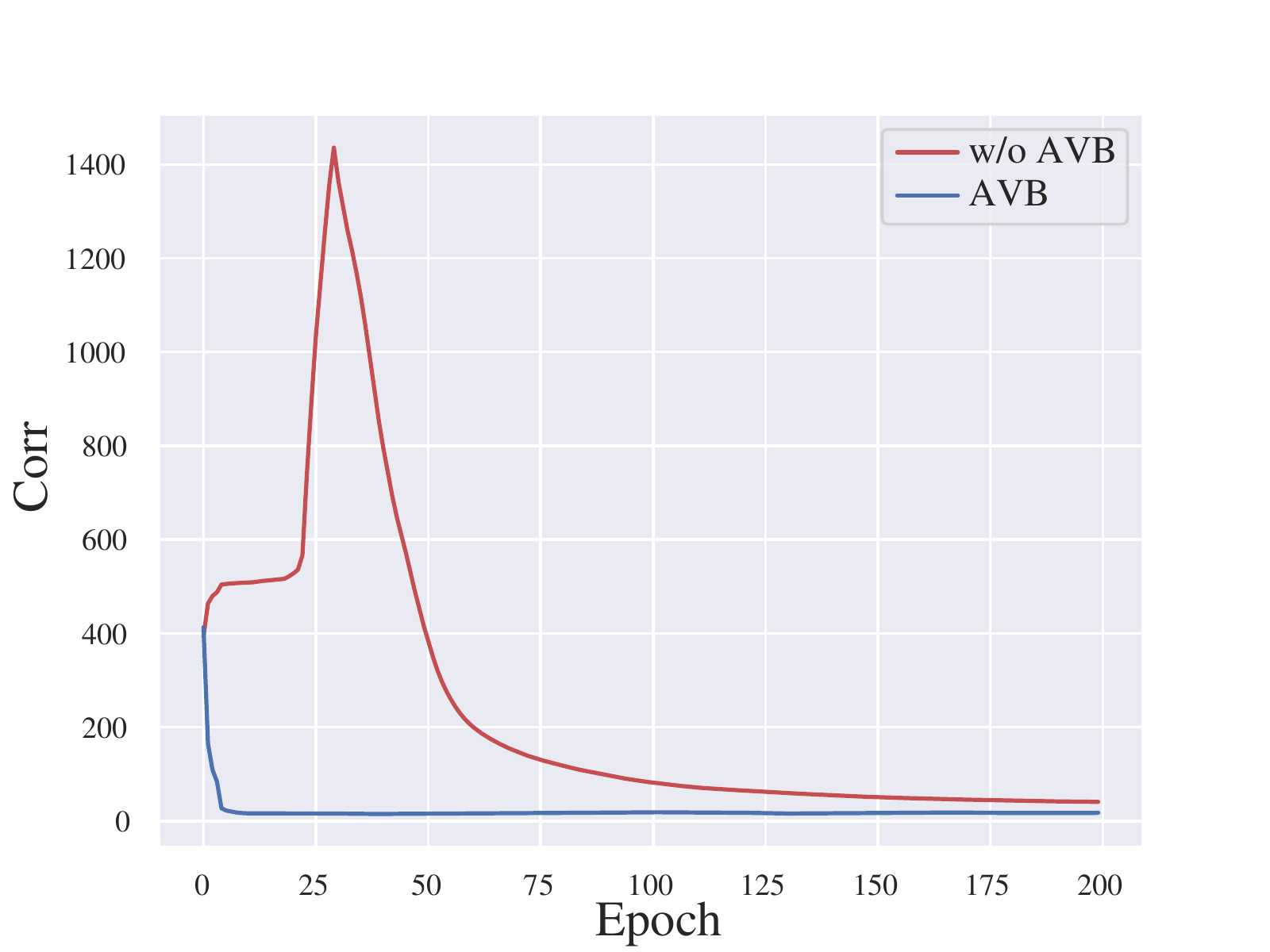}
\label{fig:corr-10m}
\end{minipage}%
}\subfigure[Yelp]{
\begin{minipage}[c]{0.24\textwidth}
\centering
\includegraphics[width=\linewidth]{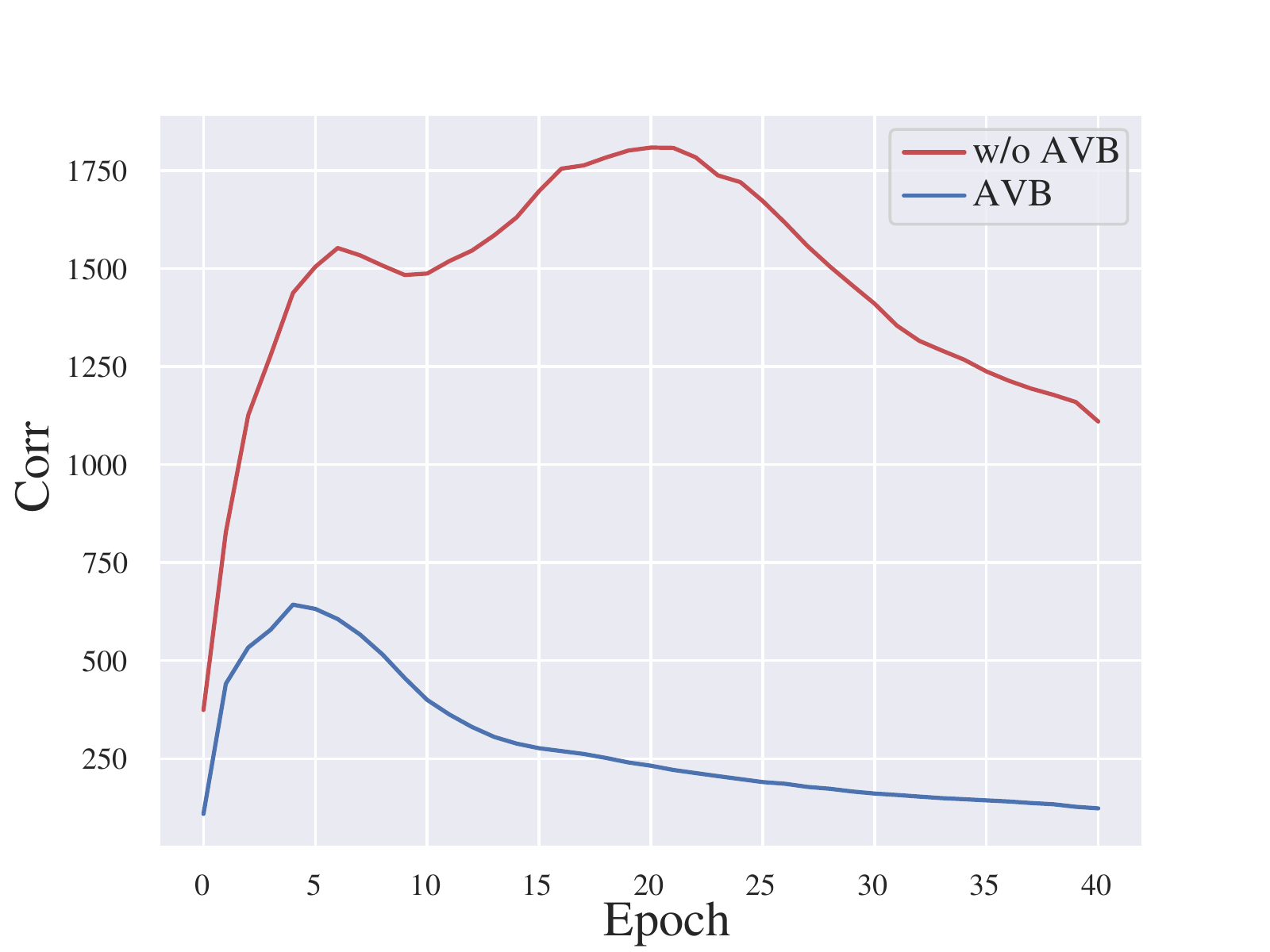}
\label{fig:corr-yelp}
\end{minipage}%
}
\caption{$corr$ on ML-latest, ML-1m, ML-10m and Yelp.}
\label{fig:corr-study}
\end{figure*}

Compared with deep learning based methods Caser and GRU4Rec$^+$, our model achieves significant improvement. On the one hand, it shows powerful predictive ability of generative models, on the other hand, it also shows that the combination of RNN and CNN may bring improvements. In CFGAN, it employs a GAN structure to generate fake purchase vectors. Although it has generative capabilities, it is not suitable for sequential recommendation, the dynamic changes of user interests can not be captured. 

Compared with BERT4Rec, although ACVAE only considers the unidirectional information, it does not utilize the global attention mechanism, it still achieves better results. It indicates that for sequence-oriented generative models, obtaining high-quality latent variables is the key to achieving good results.

It is worth noting that SVAE has achieved the best except ACVAE in most of the datasets. However, in datasets with fewer users (\emph{e.g.} ML-latest), FPMC performs better than SVAE. It shows that traditional methods are still valid on some datasets. In datasets with short average sequence length (\emph{e.g.} Yelp), BERT4Rec and Mult-VAE perform well, it shows that these two methods are competitive in sequential recommendation with short sequence length. 

\subsection{Ablation Study}
Apart from making comparison with other models, we also perform ablation study on our own model to investigate the effectiveness of different components. We choose Recall@10 as the evaluation metric. We disable the key parts of our model (without contrastive loss, without AVB, without CNN layer) in turn, and test the performance on four datasets. Figure \ref{fig:ml-latest}, \ref{fig:ml-1m}, \ref{fig:ml-10m} and \ref{fig:yelp} show the results. We can observe that training curves are smooth, and ACVAE with full components performs best, it shows the effectiveness of all the three components and each of them contributes to the result.

\subsubsection{Study of AVB}
For the model without AVB, we remove the adversary $T_{\Psi}$ without changing the structure of encoder. The results of the model without AVB is worse than the results of ACVAE in all of the metrics shown in figure \ref{fig:ablation-study}. The possible reason is that the adversary can effectively regularize the latent variables, which improves the generation ability of latent variables.

In addition, we measure the correlation coefficients of the various dimensions of the latent variables on four datasets. Since the correlation coefficients are in the form of matrix, we need to transform it into a scalar in order to evaluate the correlation in a more intuitive way. The diagonal elements in the correlation coefficient matrix are all one and the other elements represent the correlation between different elements. So we use the following formula to get a specific value to measure the correlation:
\begin{equation}
    corr = \sum_{i}\sum_{j}c_{ij}^2
\end{equation}
where $c_{ij}$ is the element of the matrix $\boldsymbol{C}-\boldsymbol{I}$ and $\boldsymbol{C}$ denotes the correlation coefficient matrix and $\boldsymbol{I}$ denotes the identity matrix.

Figure \ref{fig:corr-study} shows the change of value $corr$ on four datasets. The result shows that in all of the four datasets, the correlation coefficient of using AVB is lower than that of not using AVB. This demonstrates the important role of AVB in reducing the correlation of different dimensions of latent variables.

\subsubsection{Study of Contrastive Loss}
To verify the effectiveness of contrastive loss, we set $\beta$ to 0, which disables the contrastive loss term without affecting the training of the VAE model. From figure \ref{fig:ablation-study}, we observe that contrastive loss brings ACVAE better performance in most of the datasets. This shows that the contrastive loss can further enhance the generalization ability of the model by minimizing contrastive loss.

\subsubsection{Study of CNN Layer}
We compare the training results with and without CNN. Fully connected layers are used to replace the original CNN layer. We can find that, compared with the model without CNN layers, ACVAE achieves better results. The reason is that CNN layer helps further capture the local information of the items.

\begin{figure}
    \centering
    \centering
    \subfigure[RC@10]{
    \begin{minipage}[c]{0.23\textwidth}
    \centering
    \includegraphics[width=\linewidth]{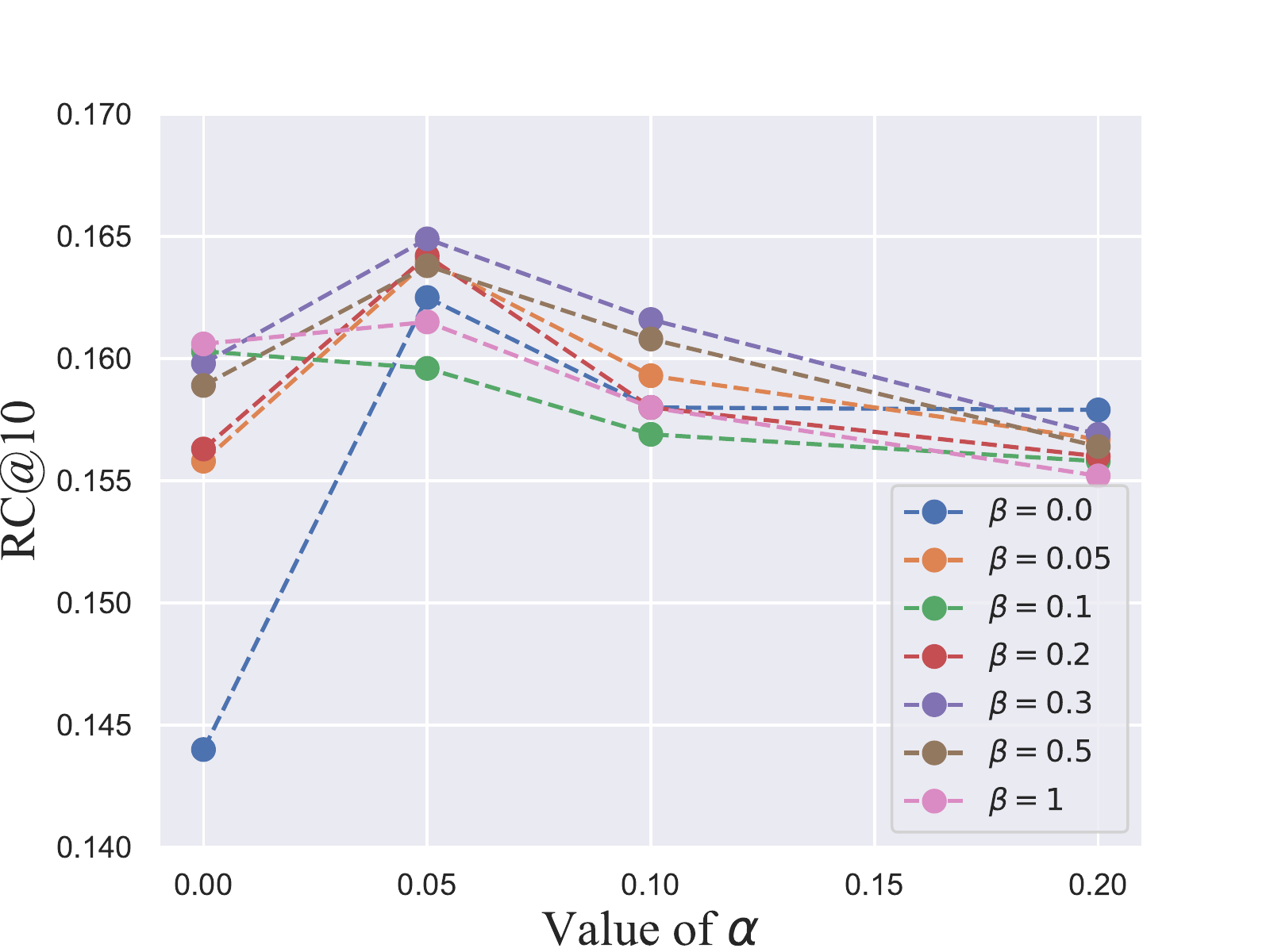}
    \label{fig:param_hr10}
    \end{minipage}%
    }
    \subfigure[NDCG@10]{
    \begin{minipage}[c]{0.23\textwidth}
    \centering
    \includegraphics[width=\linewidth]{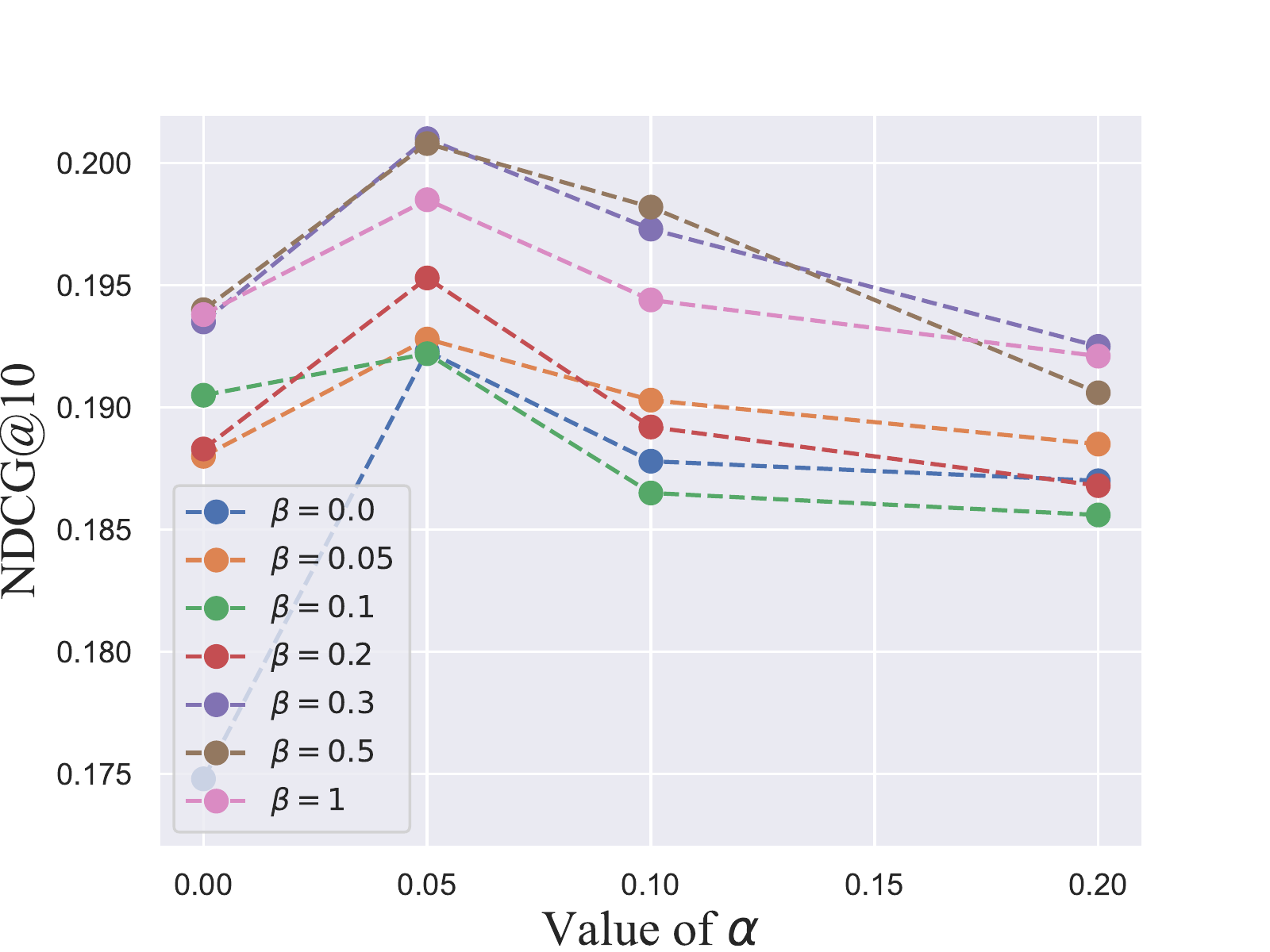}
    \label{fig:param_ndcg10}
    \end{minipage}%
    }
    \caption{Impact of parameters $\alpha$ and $\beta$ on ML-1m.}
    \label{fig:param_study}
\end{figure}

\subsection{Impact of Hyper-parameters}
In order to investigate the influence of the hyper-parameters (\emph{i.e.} $\alpha$ and $\beta$) in the objective function, we perform a grid search strategy to test the impact on ML-1M dataset. We choose Recall@10 and NDCG@10 as evaluation metrics. The results are shown in Figure \ref{fig:param_hr10} and Figure \ref{fig:param_ndcg10}.

\subsubsection{Impact of $\alpha$}
To further study the impact of the adversarial term in the objective function during training, we test the hyper-parameter $\alpha$ in the range of $\{0.0, 0.05, 0.1, 0.2\}$. According to \cite{higgins2016beta}, higher $\alpha$ values will result in stronger constraint on the latent variables. Therefore, we set the value of $\alpha$ between $0\sim0.2$ based on experience. Figure \ref{fig:param_study} shows the highest evaluation results during training of RC@10 and NDCG@10. We can find that both RC@10 and NDCG@10 reach the highest values when $\alpha=0.05$ and then go down. The reason for this result is that adversarial learning can bring certain constraints to latent variables to prevent overfitting, but too large weight of $\alpha$ will lead to over-regularization and reduce the effect.
\subsubsection{Impact of $\beta$}
To study the impact of contrastive learning, we test the hyper-parameter $\beta$ determining the weight of contrastive loss in the range of $\{0.0, 0.05, 0.1, 0.2, 0.3, 0.5, 1.0\}$. We can observe that the contrastive loss term can bring positive effect for the results, the results are best when $\beta=0.5$. It is worth noticing that when $\alpha=0.0$, using contrastive loss term will greatly improve the results, because without the KL term, the model will rely on the contrastive loss for learning.

\section{Conclusion}
In this paper, we focus on the shortcomings in the VAE models for sequential recommendation, especially the quality of the inferred latent variables. These shortcomings have largely limited the ability of latent variables in the VAE model in expressing the sequential information with users' unique preferences. We propose a novel sequential recommendation model ACVAE to enhance the encoder. We introduce adversarial learning via AVB framework to sequential recommendation, which reduces the relevance between different dimensions in latent variables. We also employ contrastive learning into VAE, which brings the model better generalization by minimizing contrastive loss. Besides, we add a special convolutional layer in the encoder after recurrent layer to further capture the short-term information in the sequence. Experiments demonstrate that our proposed ACVAE model achieves considerable performance improvement compared with state-of-the-art models.

\bibliographystyle{ACM-Reference-Format}
\bibliography{ref}
\end{document}